\documentclass[reqno]{amsart}
\usepackage{amssymb}

\newtheorem{theorem}{Theorem}[section]
\newtheorem{proposition}[theorem]{Proposition}
\newtheorem{lemma}[theorem]{Lemma}

\theoremstyle{definition}
\newtheorem*{definition}{Definition}

\theoremstyle{remark}
\newtheorem*{remark}{Remark}

\numberwithin{equation}{section}

\begin{document}

\title[Discrete Delta Bose Gas on the Circle]
{Diagonalization of an Integrable Discretization of the Repulsive
Delta Bose Gas on the Circle}

\author{J.F. van Diejen}
\address{
Instituto de Matem\'atica y F\'{\i}sica, Universidad de Talca,
Casilla 747, Talca, Chile}

\thanks{Work supported in part by the Fondo Nacional de Desarrollo
Cient\'{\i}fico y Tecnol\'ogico (FONDECYT) Grant \# 1051012, by the
Anillo Ecuaciones Asociadas a Reticulados financed by the World Bank
through the Programa Bicentenario de Ciencia y Tecnolog\'{\i}a, and
by the Programa Reticulados y Ecuaciones of the Universidad de
Talca.}


\begin{abstract}
We introduce an integrable lattice discretization of the quantum
system of $n$ bosonic particles on a ring interacting pairwise via
repulsive delta potentials. The corresponding (finite-dimensional)
spectral problem of the integrable lattice model is solved by means
of the Bethe Ansatz method. The resulting eigenfunctions turn out to
be given by specializations of the Hall-Littlewood polynomials. In
the continuum limit the solution of the repulsive delta Bose gas due
to Lieb and Liniger is recovered, including the orthogonality of the
Bethe wave functions first proved by Dorlas (extending previous work
of C.N. Yang and C.P. Yang).
\end{abstract}

\maketitle

\tableofcontents

\section{Introduction}\label{sec1}
The non-ideal Bose gas with delta-potential interactions is a
system of $n$ one-dimensional bosonic particles characterized by a
Hamiltonian given by the formal Schr\"odinger operator
\begin{equation}\label{boseham}
H=-\Delta  + g \sum_{1\leq j\neq k\leq n}
\boldsymbol{\delta}(x_j-x_k).
\end{equation}
Here $x_1,\ldots ,x_n$ represent the position variables, $\Delta
:=
\partial_{x_1}^2+\cdots +\partial_{x_n}^2$, $\boldsymbol{\delta}$
refers to the delta distribution, and $g$ denotes a coupling
parameter determining the strength of the interaction. For $g>0$
the interaction between the particles is repulsive and for $g<0$
it is attractive, whereas for $g=0$ the model degenerates to an
ideal one-dimensional boson gas without interaction between the
particles.

The eigenvalue problem for the above Schr\"odinger operator with
periodic boundary conditions---i.e. for particles moving along a
circle---was solved by Lieb and Liniger by means of the Bethe Ansatz
method \cite{lie-lin:exact}. The corresponding spectral problem for
particles moving along the whole real line was considered
subsequently by McGuire \cite{mcg:study}. Since the appearance of
these two pioneering papers, the exactly solvable quantum models
under consideration have been the subject of numerous studies; for
an overview of the vast literature and an extensive bibliography we
refer the reader to Refs.
\cite{mat:many-body,gau:fonction,kor-bog-ize:quantum,%
alb-kur:singular,sut:beautiful}.
Further generalizations of the one-dimensional quantum $n$-particle
system with delta-potential interactions can be found in Refs.
\cite{gau:boundary,gut:integrable,hec-opd:yang,die:plancherel},
where analogous quantum eigenvalue problems are studied in which the
permutation-symmetry is traded for an invariance with respect to the
action of more general reflection groups
\cite{bou:groupes,hum:reflection}, and also in Refs.
\cite{alb-fei-kur:integrability,alb-kur:singular,%
cau-cra:exact,hal-lan-pau:generalized},
where the delta-potential interaction between the particles is
replaced by more general zero-range point-like interactions
(involving combinations of $\boldsymbol{\delta}$ and
$\boldsymbol{\delta}^\prime$ type potentials)
\cite{alb-ges-hoe-hol:solvable}.

An important (and notoriously hard) problem connected with the Bethe
Ansatz method is the question of demonstrating the completeness of
the Bethe wave functions in a Hilbert space context. For the
non-ideal Bose gas on the line with a repulsive delta-potential
interaction ($g>0$), the spectrum of the Schr\"odinger operator is
purely continuous. The completeness of the Bethe wave functions was
proved for this case by Gaudin \cite{gau:bose1,gau:bose2}. For the
corresponding system in the attractive regime ($g<0$), the
completeness problem is much harder as multi-particle binding may
occur thus giving rise to mixed continuous-discrete spectrum. In
this more complex situation the completeness of the Bethe wave
functions was shown by Oxford \cite{oxf:hamiltonian}, with the aid
of techniques developed by Babbitt and Thomas in their treatment of
an analogous spectral problem for the one-dimensional infinite
isotropic Heisenberg spin chain \cite{tho:ground,bab-tho:ground}.
When passing from particles on the line to particles on the circle
the nature of the system changes drastically, as the confinement to
a compact region forces the spectrum of the Schr\"odinger operator
to become purely discrete. In this situation the completeness of the
Bethe wave functions was proved for the repulsive regime by Dorlas
\cite{dor:orthogonality}. In Dorlas' approach the question of the
completeness is first reduced to that of the orthogonality of the
Bethe wave functions. This orthogonality is then shown to hold with
the aid of quantum inverse scattering theory
\cite{kor-bog-ize:quantum}, combined with previous results of C.N.
Yang and C.P. Yang pertaining to the solution of the associated
algebraic system of Bethe equations (determining the spectrum of the
Schr\"odinger operator under consideration)
\cite{yan-yan:thermodynamics}. For the attractive regime such
progress has yet to be made: the question of the construction of a
complete eigenbasis for the non-ideal Bose gas on the circle with
delta-potential interaction remains (to date) open.

By exploiting the translational invariance and the permutation
symmetry the eigenvalue problem characterized by the Hamiltonian $H$
\eqref{boseham} reduces---in the case of bosonic particles moving
along a circle of unit circumference---to that of the free Laplacian
\begin{equation}\label{ep}
-\Delta \psi = E \psi,
\end{equation}
acting on a domain of wave functions $\psi (x_1,\ldots ,x_n)$ with
support inside the alcove
\begin{equation}\label{alcove}
\boldsymbol{A} =\{ \mathbf{x}\in\mathbb{R}^n\mid  x_1+\cdots
+x_n=0,\; x_1\geq x_2\geq \cdots\geq x_n,\; x_1-x_n\leq 1 \} ,
\end{equation}
and subject to normal linear homogeneous boundary conditions at
the walls of the alcove of the form
\begin{subequations}
\begin{eqnarray}\label{wall1}
\left. (\partial_{x_j}-\partial_{x_{j+1}}-g)\psi
\right|_{x_j-x_{j+1}=0}=0,&& j=1,\ldots, n-1 , \\
 \left.
(\partial_{x_n}-\partial_{x_{1}}-g)\psi \right|_{x_1-x_{n}=1}=0 .&&
\label{wall2}
\end{eqnarray}
\end{subequations}
(The parameter $E$ represents the energy eigenvalue.) The purpose of
the present paper is to study a discretization of the eigenvalue
problem in Eqs. \eqref{ep}--\eqref{wall2}. Throughout we will
restrict attention to the repulsive parameter regime $g>0$.

More specifically, we study the eigenvalue problem for an integrable
system of discrete Laplacians acting on functions supported on a
regular lattice over the alcove $\boldsymbol{A}$ \eqref{alcove}, and
subject to repulsive reflection relations at the boundary of the
lattice. Since the alcove is compact, the lattice in question is
finite; hence, our discrete eigenvalue problem is
finite-dimensional. We solve the eigenvalue problem at issue by
means of the Bethe Ansatz method. The resulting Bethe eigenfunctions
turn out to be given by specializations of the Hall-Littlewood
polynomials \cite{mac:symmetric,mac:orthogonal}. The orthogonality
and completeness of these Bethe eigenfunctions arises as an
immediate consequence of the integrability (which permits removing
possible degeneracies in the spectrum of the Laplacians). As a
byproduct, the Lieb-Liniger type Bethe eigenfunctions for the
eigenvalue problem in Eqs. \eqref{ep}--\eqref{wall2} are recovered
via a continuum limit. The orthogonality of the latter
eigenfunctions (and thus eventually---because of Dorlas' results
\cite{dor:orthogonality}---also the completeness) are in our
approach immediately inherited from the corresponding orthogonality
results for our discretized lattice model. In this connection it is
probably helpful to recall that the original proof of the
orthogonality due to Dorlas \cite{dor:orthogonality} also involves a
discretization, which arises however in a fundamentally different
way from the one employed here. In a nutshell: Dorlas arrives at the
orthogonality through a continuum limit of the (second) quantization
of the Lattice Nonlinear Schr\"odinger Equation introduced by
Izergin and Korepin \cite{kor-bog-ize:quantum}, whereas here---in
contrast---we study a rather more elementary quantum lattice model
characterized by a direct discretization of the Schr\"odinger
operator in Eqs. \eqref{ep}--\eqref{wall2} itself.

The paper is structured as follows. In Section \ref{sec2} the
discretization of the eigenvalue problem in Eqs.
\eqref{ep}--\eqref{wall2} is formulated. In Section \ref{sec3} the
eigenfunctions are constructed by means of the Bethe Ansatz method.
The associated Bethe equations are solved in Section \ref{sec4} and
the orthogonality and completeness of the corresponding Bethe wave
functions is demonstrated in Section \ref{sec5}. Finally, the
continuum limit is analyzed in Section \ref{sec6}.

\section{Discrete Laplacians on the Alcove}\label{sec2}
In this section we introduce a system of discrete Laplacians on a
finite lattice over (a dilated version of) the alcove
$\boldsymbol{A}$ \eqref{alcove}. For this purpose it will be
convenient to borrow concepts and notation from the theory of root
systems. Here we will only need to deal with the simplest type of
root systems: those of type $A$. For further background material
concerning root systems the reader is referred to Refs.
\cite{bou:groupes,hum:reflection}.

\subsection{Preliminaries}
Let $\mathbf{e}_1,\ldots ,\mathbf{e}_n$ denote the standard basis of
unit vectors in $\mathbb{R}^n$ and let $\langle \cdot ,\cdot\rangle$
be the (usual) inner product with respect to which the standard
basis is orthonormal. The alcove $\boldsymbol{A}$ \eqref{alcove}
constitutes a convex polyhedron in the center-of-mass plane
\begin{equation}\label{mass-plane} \boldsymbol{E}=\{
\mathbf{x}\in \mathbb{R}^n \mid \langle
\mathbf{x},\mathbf{e}\rangle =0\} ,\quad
\mathbf{e}=\mathbf{e}_1+\cdots +\mathbf{e}_n,
\end{equation}
which is bounded by the $n$ hyperplanes
\begin{subequations}
\begin{eqnarray}
&& \boldsymbol{E}_0=\{ \mathbf{x}\in\boldsymbol{E}\mid \langle
\mathbf{x},\boldsymbol{\alpha}_0\rangle =1 \} ,\label{pl1} \\
 &&\boldsymbol{E}_j=\{ \mathbf{x}\in\boldsymbol{E}\mid \langle
\mathbf{x},\boldsymbol{\alpha}_j\rangle =0 \} ,\quad j=1,\ldots ,
n-1,\label{pl2}
\end{eqnarray}
\end{subequations}
where
\begin{equation}\label{sroot}
\boldsymbol{\alpha}_0:=\mathbf{e}_1-\mathbf{e}_{n} \quad
\text{and}\quad
\boldsymbol{\alpha}_j:=\mathbf{e}_j-\mathbf{e}_{j+1},\quad
j=1,\ldots ,n-1 .
\end{equation}
Specifically, we have that
\begin{equation}\label{alcove2}
\boldsymbol{A}=\{\mathbf{x}\in\boldsymbol{E}\mid \langle
\mathbf{x},\boldsymbol{\alpha}_0\rangle \leq 1 ,\; \langle
\mathbf{x},\boldsymbol{\alpha}_j\rangle \geq 0,\, j=1,\ldots ,n-1 \}
.
\end{equation}
The vertices (i.e. corners) of the polyhedron $\boldsymbol{A}$ are
determined by the intersections of all choices of $n-1$($=\dim
(\boldsymbol{E})$) out of the $n$ hyperplanes
$\boldsymbol{E}_0,\ldots ,\boldsymbol{E}_{n-1}$. These vertices are
given explicitly by the origin $\boldsymbol{0}$ and the vectors
\begin{equation}\label{fweight}
\boldsymbol{\omega}_j:=\mathbf{e}_1+\cdots +\mathbf{e}_j-{\textstyle
\frac{j}{n}}(\mathbf{e}_{1}+\cdots +\mathbf{e}_n),\qquad j=1,\ldots
,n-1 .
\end{equation}
Indeed, the vectors $\boldsymbol{\omega}_1,\ldots
,\boldsymbol{\omega}_{n-1}$ all lie on the hyperplane
$\boldsymbol{E}_0$ \eqref{pl1} and constitute a basis of
$\boldsymbol{E}$ that is dual to the basis
$\boldsymbol{\alpha}_1,\ldots ,\boldsymbol{\alpha}_{n-1}$ (in the
sense that $\langle \boldsymbol{\omega}_j
,\boldsymbol{\alpha}_k\rangle =\delta_{j,k}$, where $\delta_{j,k}$
denotes the Kronecker delta symbol).

Let $r_0:\boldsymbol{E}\to\boldsymbol{E}$ be the orthogonal
reflection in the hyperplane $ \boldsymbol{E}_0$ \eqref{pl1} and let
$r_j:\boldsymbol{E}\to\boldsymbol{E}$, $j=1,\ldots ,n-1$ be the
orthogonal reflections in the hyperplanes $\boldsymbol{E}_j$
\eqref{pl2}. The action of these reflections on an arbitrary vector
$\mathbf{x}\in \boldsymbol{E}$ is of the form
\begin{subequations}
\begin{eqnarray}\label{r0}
r_0(\mathbf{x})&=&\mathbf{x}+(1-\langle
\mathbf{x},\boldsymbol{\alpha}_0\rangle )\boldsymbol{\alpha}_0, \\
\label{rj} r_j(\mathbf{x})&=&\mathbf{x}-\langle
\mathbf{x},\boldsymbol{\alpha}_j\rangle \boldsymbol{\alpha}_j, \quad
j=1,\ldots ,n-1.
\end{eqnarray}
\end{subequations}
From these two formulas it is readily inferred that for $j\in \{
1,\ldots ,n-1\}$ the reflection $r_j$ swaps the $j^{th}$ and
$(j+1)^{th}$ coordinates of $\mathbf{x}$, and that $r_0$ swaps the
first and the last coordinates followed by a translation over the
vector $\boldsymbol{\alpha}_0$. Hence, the reflections $r_1,\ldots
,r_{n-1}$ generate an action of the permutation group
$\mathcal{S}_n$ on $\boldsymbol{E}$, and the reflections $r_0,\ldots
,r_{n-1}$ generate and action of the affine permutation group
$\hat{\mathcal{S}}_n=\mathcal{S}_n\ltimes \mathcal{Q}$, which is the
semidirect product of the permutation group $\mathcal{S}_n$ and the
lattice of translations
\begin{equation}\label{rootlat}
\mathcal{Q}:= \text{Span}_{\mathbb{Z}}(\boldsymbol{\alpha}_1,\ldots
,\boldsymbol{\alpha}_{n-1}) .
\end{equation}
For any (affine) permutation $\sigma \in \hat{\mathcal{S}}_n$, one
defines its {\em length} $\ell (\sigma)$ as the minimal number of
reflections needed for decomposing $\sigma$ (non-uniquely) in terms
of the generators:
\begin{equation}
\sigma = r_{j_1}r_{j_2}\cdots r_{j_\ell}
\end{equation}
(where $j_1,\ldots ,j_\ell\in \{ 0,1,\ldots ,n-1\}$ and with the
convention that the length of the identity element is equal to
zero).

The polyhedron $\boldsymbol{A}$ \eqref{alcove2} constitutes a
fundamental domain for the action of $\hat{\mathcal{S}}_n$ on
$\boldsymbol{E}$. More specifically, for each $\mathbf{x}\in
\boldsymbol{E}$ the orbit $\hat{\mathcal{S}}_n(\mathbf{x})$
intersects $\boldsymbol{A}$ precisely once. Let us denote by
$\sigma_{\mathbf{x}}\in \hat{\mathcal{S}}_n$ the unique shortest
affine permutation such that
\begin{equation}
\sigma_{\mathbf{x}}(\mathbf{x})\in \boldsymbol{A}.
\end{equation}

Let us fix a positive integer $m$.  Below it will often be
convenient to employ a dilated version of the polyhedron
$\boldsymbol{A}^{(m)}:=m\boldsymbol{A}$ rather than $\boldsymbol{A}$
itself. Throughout we shall distinguish by means of a superscript
$(m)$ the corresponding boundary planes, boundary reflections, and
the elements of the affine permutation group
$\hat{\mathcal{S}}_n^{(m)}:=\mathcal{S}_n\ltimes
(m\mathcal{Q})\subset \hat{\mathcal{S}}_n$ generated by these
reflections. That is to say, the dilated alcove
$\boldsymbol{A}^{(m)}$ is bounded by the hyperplanes
$\boldsymbol{E}^{(m)}_j:=m\boldsymbol{E}_j$, $j=0,\ldots ,n-1$; the
orthogonal reflections in these hyperplanes act as
$r_j^{(m)}(\mathbf{x}):=mr_j(\mathbf{x}/m)$, $j=0,\ldots ,n-1$.  (So
$\boldsymbol{E}^{(m)}_j=\boldsymbol{E}_j$ and $r_j^{(m)}=r_j$ if
$j>0$.) The affine permutation
$\sigma^{(m)}_{\mathbf{x}}\in\hat{\mathcal{S}}_n^{(m)}$ mapping a
vector $\mathbf{x}\in\boldsymbol{E}$ into the fundamental domain
$\boldsymbol{A}^{(m)}$ is given by the action
$\sigma^{(m)}_{\mathbf{x}}(\mathbf{x}):= m
\sigma_{\mathbf{x}/m}(\mathbf{x}/m)$.

\subsection{Laplacians}
The orthogonal projection of $\mathbb{Z}^n\subset\mathbb{R}^n$ onto
the center-of-mass plane $\boldsymbol{E}$ \eqref{mass-plane} is
given by the lattice dual to $\mathcal{Q}$ \eqref{rootlat}:
\begin{subequations}
\begin{eqnarray}\label{weightlat1}
\mathcal{P} &:=& \{ \boldsymbol{\lambda}\in\boldsymbol{E} \mid
\forall\boldsymbol{\alpha}\in\mathcal{Q}:\; \langle
\boldsymbol{\lambda},\boldsymbol{\alpha}\rangle\in\mathbb{Z}\}  \\
 &=&
\text{Span}_{\mathbb{Z}}(\boldsymbol{\omega}_1,\ldots
,\boldsymbol{\omega}_{n-1}) . \label{weightlat2}
\end{eqnarray}
\end{subequations}
It is clear from Eq. \eqref{weightlat1} that $\mathcal{Q}$ is
contained as a sublattice in $\mathcal{P}$. Hence, the action of the
affine permutation group $\hat{\mathcal{S}}_n$ in $\boldsymbol{E}$
maps the lattice $\mathcal{P}$ into itself (cf. Eqs. \eqref{r0},
\eqref{rj}). The intersection of the lattice $\mathcal{P}$ with the
dilated polyhedron $\boldsymbol{A}^{(m)}$ provides a finite grid
$\mathcal{P}^{(m)}$ over $\boldsymbol{A}^{(m)}$ containing its
vertices $\mathbf{0}$ and $m\omega_1,\ldots ,m\omega_{n-1}$:
\begin{eqnarray}
\label{pm} \mathcal{P}^{(m)} &:=& \{ \boldsymbol{\lambda} \in
\mathcal{P} \mid \boldsymbol{\lambda}\in
\boldsymbol{A}^{(m)}\} \\
&=& \{ k_1\omega_1+\cdots k_{n-1}\omega_{n-1} \mid k_1,\ldots
,k_{n-1}\in\mathbb{Z}_{\geq 0},\; k_1+\cdots +k_{n-1}\leq m\} .
\nonumber
\end{eqnarray}
We are now in the position to define a system of $n-1$ Laplace
operators acting in the space $\mathcal{C}(\mathcal{P}^{(m)})$ of
complex functions $\psi:\mathcal{P}^{(m)}\to\mathbb{C}$.

\begin{definition}[Laplace Operator]
To each basis vector $\boldsymbol{\omega}_k$ from Eq.
\eqref{fweight} we associate a corresponding Laplace operator
$L^{(m)}_k:\mathcal{C}(\mathcal{P}^{(m)})\to\mathcal{C}(\mathcal{P}^{(m)})$
defined by its action on an arbitrary function
$\psi:\mathcal{P}^{(m)}\to\mathbb{C}$ of the form
\begin{subequations}
\begin{equation}\label{lapk}
(L^{(m)}_k\psi)_{\boldsymbol{\lambda}} := \sum_{\boldsymbol{\nu}\in
\mathcal{S}_n(\boldsymbol{\omega}_k)}
\psi_{\boldsymbol{\lambda}+\boldsymbol{\nu}} ,
\end{equation}
with the boundary convention that for $\boldsymbol{\mu}\in
\mathcal{P}\setminus\mathcal{P}^{(m)}$
\begin{equation}\label{bconv}
\psi_{\boldsymbol{\mu}}:=t^{\ell^{(m)} (
\sigma^{(m)}_{\boldsymbol{\mu}})}
\psi_{\sigma^{(m)}_{\boldsymbol{\mu}}(\boldsymbol{\mu})}
,
\end{equation}
\end{subequations}
where $t$ denotes a real (coupling) parameter (and the length
function $\ell^{(m)}(\cdot )$ refers to the minimal number of
reflections in the decomposition of an affine permutation in
$\hat{\mathcal{S}}_n^{(m)}$ in terms of $r^{(m)}_0,\ldots
,r^{(m)}_{n-1}$).
\end{definition}

Roughly speaking, the value of $L^{(m)}_k\psi$ in a point
$\boldsymbol{\lambda}\in\mathcal{P}^{(m)}$ is equal to the sum of
the values of $\psi$ in all neighboring points of the form
$\boldsymbol{\lambda}+\boldsymbol{\nu}$, where $\boldsymbol{\nu}$
runs through the orbit of $\boldsymbol{\omega}_k$ with respect to
the action of the permutation group $\mathcal{S}_n$. When
$\boldsymbol{\lambda}+\boldsymbol{\nu}$ lies outside
$\mathcal{P}^{(m)}$ the value of
$\psi_{\boldsymbol{\lambda}+\boldsymbol{\nu}}$ is governed by the
boundary convention in Eq. \eqref{bconv}. It is instructive to
clarify the nature of this boundary convention somewhat more in
detail by decomposing
$\sigma^{(m)}_{\boldsymbol{\lambda}+\boldsymbol{\nu}}$ in terms of
the elementary reflections in the hyperplanes bounding
$\boldsymbol{A}^{(m)}$.

\begin{proposition}[Boundary Reflection Relations]
\label{bref:prp} Let $\boldsymbol{\lambda}\in\mathcal{P}^{(m)}$. The
boundary convention in Eq. \eqref{bconv} amounts to the requirement
that $\forall \boldsymbol{\nu}\in
\mathcal{S}_n(\boldsymbol{\omega}_k)$ for which
$\boldsymbol{\lambda}+\boldsymbol{\nu}\in
\mathcal{P}\setminus\mathcal{P}^{(m)}$
\begin{subequations}
\begin{equation}\label{bref1}
\psi_{\boldsymbol{\lambda}+\boldsymbol{\nu}}=
\begin{cases}
t \psi_{r_0^{(m)}(\boldsymbol{\lambda}+\boldsymbol{\nu})}
&\text{if}\
\langle\boldsymbol{\lambda}+\boldsymbol{\nu},
\boldsymbol{\alpha}_0\rangle>m \\
t \psi_{r_j^{(m)}(\boldsymbol{\lambda}+\boldsymbol{\nu})}
&\text{if}\ \langle\boldsymbol{\lambda}+\boldsymbol{\nu},
\boldsymbol{\alpha}_j\rangle<0\ \ (j>0),
\end{cases}
\end{equation}
or equivalently
\begin{equation}\label{bref2}
\psi_{\boldsymbol{\lambda}+\boldsymbol{\nu}}=
\begin{cases}
t \psi_{\boldsymbol{\lambda}+\boldsymbol{\nu}-\boldsymbol{\alpha}_0}
&\text{if}\
\langle\boldsymbol{\lambda},\boldsymbol{\alpha}_0\rangle=m\;\text{and}\;
\langle\boldsymbol{\nu},\boldsymbol{\alpha}_0\rangle=1 \\
t \psi_{\boldsymbol{\lambda}+\boldsymbol{\nu}+\boldsymbol{\alpha}_j}
&\text{if}\
\langle\boldsymbol{\lambda},\boldsymbol{\alpha}_j\rangle=0\;\text{and}\;
\langle\boldsymbol{\nu},\boldsymbol{\alpha}_j\rangle=-1\ (j>0).
\end{cases}
\end{equation}
\end{subequations}
\end{proposition}
\begin{proof}
Let $\boldsymbol{\lambda}\in \mathcal{P}^{(m)}$ and
$\boldsymbol{\nu}\in \mathcal{S}_n(\boldsymbol{\omega}_k)$ such that
$\boldsymbol{\lambda}+\boldsymbol{\nu}\in \mathcal{P}\setminus
\mathcal{P}^{(m)}$. Then there exist $j\in \{0,\ldots ,n-1\}$ such
that $\langle \boldsymbol{\lambda}+\boldsymbol{\nu}
,\boldsymbol{\alpha}_j\rangle $ is either $>m$ if $j=0$ or $<0$ if
$j>0$. Geometrically, this means that the hyperplane
$\boldsymbol{E}^{(m)}_j$ separates
$\boldsymbol{\lambda}+\boldsymbol{\nu}$ from
$\sigma^{(m)}_{\boldsymbol{\lambda}+\boldsymbol{\nu}}
(\boldsymbol{\lambda}+\boldsymbol{\nu})\in\mathcal{P}^{(m)}$. Let us
write
$\boldsymbol{\mu}:=r^{(m)}_j(\boldsymbol{\lambda}+\boldsymbol{\nu})$.
Then for any such $j$ we have that
\begin{equation*}
\sigma^{(m)}_{\boldsymbol{\lambda}+\boldsymbol{\nu}}=
\sigma^{(m)}_{\boldsymbol{\mu}}r^{(m)}_{j} \quad\text{with}\quad
\ell^{(m)} (\sigma^{(m)}_{\boldsymbol{\mu}})= \ell^{(m)}
(\sigma^{(m)}_{\boldsymbol{\lambda}+\boldsymbol{\nu}})-1 .
\end{equation*}
We will now use induction on the length of
$\sigma^{(m)}_{\boldsymbol{\lambda}+\boldsymbol{\nu}}$ to prove that
the boundary convention in Eq. \eqref{bconv} and the boundary
reflection relation in Eq. \eqref{bref1} are equivalent. Indeed,
upon assuming \eqref{bconv} it is clear that
\begin{equation*}
\psi_{\boldsymbol{\lambda}+\boldsymbol{\nu}}= t^{\ell^{(m)} (
\sigma^{(m)}_{\boldsymbol{\lambda}+\boldsymbol{\nu}})}
\psi_{\sigma^{(m)}_{\boldsymbol{\lambda}+
\boldsymbol{\nu}}(\boldsymbol{\lambda}+\boldsymbol{\nu})}
=  t^{\ell^{(m)} ( \sigma^{(m)}_{\boldsymbol{\mu}})+1}\,
\psi_{\sigma^{(m)}_{\boldsymbol{\mu}}(\boldsymbol{\mu})}
=t\,\psi_{\boldsymbol{\mu}} ,
\end{equation*}
which amounts to \eqref{bref1}. Reversely, upon assuming
\eqref{bref1} and invoking the induction hypothesis we see that
\begin{equation*}
\psi_{\boldsymbol{\lambda}+\boldsymbol{\nu}}=
t\,\psi_{\boldsymbol{\mu}}=  t^{\ell^{(m)} (
\sigma^{(m)}_{\boldsymbol{\mu}})+1}\,
\psi_{\sigma^{(m)}_{\boldsymbol{\mu}}(\boldsymbol{\mu})} =
t^{\ell^{(m)} (
\sigma^{(m)}_{\boldsymbol{\lambda}+\boldsymbol{\nu}})}
\psi_{\sigma^{(m)}_{\boldsymbol{\lambda}+
\boldsymbol{\nu}}(\boldsymbol{\lambda}+\boldsymbol{\nu})}
,
\end{equation*}
which amounts to \eqref{bconv}. To finish the proof of the
proposition it remains to check that the boundary reflection
relations in Eqs. \eqref{bref1} and \eqref{bref2} are equivalent.
For this purpose it suffices to notice that the requirements that
$\boldsymbol{\lambda}\in\mathcal{P}^{(m)}$ and
$\boldsymbol{\nu}\in\mathcal{S}_n(\boldsymbol{\omega_k})$ imply that
$0\leq \langle \boldsymbol{\lambda},\boldsymbol{\alpha}_j\rangle
\leq m$ and that $-1\leq \langle
\boldsymbol{\nu},\boldsymbol{\alpha}_j\rangle \leq 1$ for
$j=0,\ldots ,n-1$. Hence
$\langle\boldsymbol{\lambda}+\boldsymbol{\nu},\boldsymbol{\alpha}_0\rangle>m$
iff $\langle\boldsymbol{\lambda},\boldsymbol{\alpha}_0\rangle=m$ and
$\langle\boldsymbol{\nu},\boldsymbol{\alpha}_0\rangle=1$, in which
case
$r_0^{(m)}(\boldsymbol{\lambda}+\boldsymbol{\nu})=
\boldsymbol{\lambda}+\boldsymbol{\nu}-\boldsymbol{\alpha}_0$,
and furthermore for $j>0$ one has that
$\langle\boldsymbol{\lambda}+\boldsymbol{\nu},\boldsymbol{\alpha}_j\rangle<0$
iff $\langle\boldsymbol{\lambda},\boldsymbol{\alpha}_j\rangle=0$ and
$\langle\boldsymbol{\nu},\boldsymbol{\alpha}_j\rangle=-1$, in which
case $r_j^{(m)}(\boldsymbol{\lambda}+\boldsymbol{\nu})=
\boldsymbol{\lambda}+\boldsymbol{\nu}+\boldsymbol{\alpha}_j$.
\end{proof}

It is clear from the proposition that the boundary convention in Eq.
\eqref{bconv} amounts to a normal linear boundary condition at the
hyperplanes bounding $\boldsymbol{A}^{(m)}$. The coupling parameter
$t$ determines the nature of this boundary condition; for $|t|>1$
the boundary term (i.e. the interaction between the particles) is
attractive whereas for $|t|<1$ it is repulsive. For $t=0$ and $t=1$
we are dealing with Dirichlet type and Neumann type boundary
conditions, respectively.

Applying the boundary convention to those contributions in the sum
over the translated orbit
$\boldsymbol{\lambda}+\mathcal{S}_n(\boldsymbol{\omega}_k)$
corresponding to lattice points outside the grid $\mathcal{P}^{(m)}$
gives rise to a closed formula for the action of the Laplace
operator in which the value of $L^{(m)}_k\psi$ at the point
$\boldsymbol{\lambda}$ is expressed completely in terms of the
values of $\psi$ at the neighboring points of the form
$\boldsymbol{\lambda}+\boldsymbol{\nu}\in\mathcal{P}^{(m)}$. To make
this procedure completely explicit we shall need some further
notation. Let $\mathbf{R}$ be the orbit of the basis
$\boldsymbol{\alpha}_1,\ldots,\boldsymbol{\alpha}_{n-1}$ with
respect to the action of the permutation group $\mathcal{S}_n$ and
let $\mathbf{R}^+$ be the part of the orbit that expands
nonnegatively with respect to this basis:
\begin{equation}
\mathbf{R}=\{ \mathbf{e}_j-\mathbf{e}_k\mid 1\leq j\neq k\leq n\} ,
\qquad \mathbf{R}^+=\{ \mathbf{e}_j-\mathbf{e}_k\mid 1\leq j< k\leq
n\} .
\end{equation}

\begin{proposition}[Explicit Action of the Laplace Operator]
\label{action:prp}
The action of the Laplace operator
$L^{(m)}_k$ ($k\in \{ 1,\ldots ,n-1\}$) on an arbitrary grid
function $\psi:\mathcal{P}^{(m)}\to\mathbb{C}$ is of the form
\begin{subequations}
\begin{equation}
(L^{(m)}_k\psi)_{\boldsymbol{\lambda}} = \sum_{\begin{subarray}{c}
\boldsymbol{\nu}\in
\mathcal{S}_n(\boldsymbol{\omega}_k)\\
\boldsymbol{\lambda}+\boldsymbol{\nu}\in\mathcal{P}^{(m)}\end{subarray}}
V^{(m)}_{\boldsymbol{\lambda},\boldsymbol{\nu}}\,
\psi_{\boldsymbol{\lambda}+\boldsymbol{\nu}}
,
\end{equation}
where
\begin{equation}\label{coefV}
V^{(m)}_{\boldsymbol{\lambda},\boldsymbol{\nu}}:=
\prod_{\begin{subarray}{c}\boldsymbol{\alpha}\in \mathbf{R}^+\\
\langle \boldsymbol{\lambda},\boldsymbol{\alpha}\rangle =0\\
\langle \boldsymbol{\nu},\boldsymbol{\alpha}\rangle =1
\end{subarray}}
\frac{1-t^{1+\langle \boldsymbol{\rho},\boldsymbol{\alpha}
\rangle}}{1-t^{\langle \boldsymbol{\rho},\boldsymbol{\alpha}
\rangle}}
\prod_{\begin{subarray}{c}\boldsymbol{\alpha}\in \mathbf{R}^+\\
\langle \boldsymbol{\lambda},\boldsymbol{\alpha}\rangle =m\\
\langle \boldsymbol{\nu},\boldsymbol{\alpha}\rangle =-1
\end{subarray}}
\frac{1-t^{1+n-\langle \boldsymbol{\rho},\boldsymbol{\alpha}
\rangle}}{1-t^{n-\langle \boldsymbol{\rho},\boldsymbol{\alpha}
\rangle}},
\end{equation}
\end{subequations}
with
$\boldsymbol{\rho}:=
\sum_{\boldsymbol{\alpha}\in\mathbf{R}^+}\boldsymbol{\alpha}/2=
\boldsymbol{\omega}_1+\cdots +\boldsymbol{\omega}_{n-1}$.
\end{proposition}
\begin{proof}
It follows from (the proof of) Proposition \ref{bref:prp} that for
$\boldsymbol{\lambda}\in\mathcal{P}^{(m)}$ and
$\boldsymbol{\nu}\in\mathcal{S}_n(\boldsymbol{\omega}_k)$ such that
$\boldsymbol{\lambda}+\boldsymbol{\nu}\in\mathcal{P}^{(m)}$ the
coefficient $V^{(m)}_{\boldsymbol{\lambda},\boldsymbol{\nu}}$  of
$\psi_{\boldsymbol{\lambda}+\boldsymbol{\nu}}$ in
$(L^{(m)}_k\psi)_{\boldsymbol{\lambda}}$ is of the form $\sum_\sigma
t^{\ell^{(m)}(\sigma)}$, where the sum is over those affine
permutations $\sigma$ in $\hat{\mathcal{S}}_n^{(m)}$ that are
generated by iterated application of reflections of the type
figuring in Eqs. \eqref{bref1}, \eqref{bref2}. These affine
permutations are precisely the $\sigma\in \hat{\mathcal{S}}_n^{(m)}$
for which $\sigma (\boldsymbol{\lambda})=\boldsymbol{\lambda}$ and
 $\sigma
(\boldsymbol{\lambda}+\boldsymbol{\nu})\not\in\mathcal{P}^{(m)}$, or
equivalently, $\sigma
(\boldsymbol{\lambda}+\boldsymbol{\nu})\neq\boldsymbol{\lambda}+\boldsymbol{\nu}$.
In other words, the coefficient is equal to the Poincar\'e series of
the quotient of the stabilizer subgroup
$\hat{\mathcal{S}}_{n,\boldsymbol{\lambda}}^{(m)}:=\{ \sigma \in
\hat{\mathcal{S}}_n^{(m)}\mid
\sigma(\boldsymbol{\lambda})=\boldsymbol{\lambda}\}$ and the
stabilizer subgroup
$\hat{\mathcal{S}}_{n,\boldsymbol{\lambda}}^{(m)}\cap
\hat{\mathcal{S}}_{n,\boldsymbol{\lambda}+\boldsymbol{\nu}}^{(m)}$,
i.e.
\begin{equation}\label{ps}
V^{(m)}_{\boldsymbol{\lambda},\boldsymbol{\nu}} = \sum_{\sigma \in
\hat{\mathcal{S}}_{n,\boldsymbol{\lambda}}^{(m)}/
(\hat{\mathcal{S}}_{n,\boldsymbol{\lambda}}^{(m)}\cap
\hat{\mathcal{S}}_{n,\boldsymbol{\lambda}+\boldsymbol{\nu}}^{(m)})}
t^{\ell^{(m)}(\sigma)} =\frac{\sum_{\sigma \in
\hat{\mathcal{S}}_{n,\boldsymbol{\lambda}}^{(m)}}
t^{\ell^{(m)}(\sigma)} }{ \sum_{\sigma \in
\hat{\mathcal{S}}_{n,\boldsymbol{\lambda}}^{(m)}\cap
\hat{\mathcal{S}}_{n,\boldsymbol{\lambda}+\boldsymbol{\nu}}^{(m)}}
t^{\ell^{(m)}(\sigma)} } .
\end{equation}
It follows from a general formula for the Poincar\'e series of
(affine) Weyl groups due to Macdonald \cite{mac:poincare} (cf.
Corollaries (2.5) and (3.4)) that the Poincar\'e series in the
numerator and denominator of Eq. \eqref{ps} admit product
representations given by
\begin{equation}\label{ps1}
\sum_{\sigma \in \hat{\mathcal{S}}_{n,\boldsymbol{\lambda}}^{(m)}}
t^{\ell^{(m)}(\sigma)} =\prod_{\begin{subarray}{c}\boldsymbol{\alpha}\in \mathbf{R}^+\\
\langle \boldsymbol{\lambda},\boldsymbol{\alpha}\rangle =0
\end{subarray}}
\frac{1-t^{1+\langle \boldsymbol{\rho},\boldsymbol{\alpha}
\rangle}}{1-t^{\langle \boldsymbol{\rho},\boldsymbol{\alpha}
\rangle}}
\prod_{\begin{subarray}{c}\boldsymbol{\alpha}\in \mathbf{R}^+\\
\langle \boldsymbol{\lambda},\boldsymbol{\alpha}\rangle =m
\end{subarray}}
\frac{1-t^{1+n-\langle \boldsymbol{\rho},\boldsymbol{\alpha}
\rangle}}{1-t^{n-\langle \boldsymbol{\rho},\boldsymbol{\alpha}
\rangle}}
\end{equation}
and
\begin{equation}\label{ps2}
\sum_{\sigma \in
\hat{\mathcal{S}}_{n,\boldsymbol{\lambda}}^{(m)}\cap
\hat{\mathcal{S}}_{n,\boldsymbol{\lambda}+\boldsymbol{\nu}}^{(m)}}
t^{\ell^{(m)}(\sigma)}= \prod_{\begin{subarray}{c}\boldsymbol{\alpha}\in \mathbf{R}^+\\
\langle \boldsymbol{\lambda},\boldsymbol{\alpha}\rangle =0\\
\langle \boldsymbol{\lambda}+
\boldsymbol{\nu},\boldsymbol{\alpha}\rangle =0
\end{subarray}}
\frac{1-t^{1+\langle \boldsymbol{\rho},\boldsymbol{\alpha}
\rangle}}{1-t^{\langle \boldsymbol{\rho},\boldsymbol{\alpha}
\rangle}}
\prod_{\begin{subarray}{c}\boldsymbol{\alpha}\in \mathbf{R}^+\\
\langle \boldsymbol{\lambda},\boldsymbol{\alpha}\rangle =m\\
\langle
\boldsymbol{\lambda}+\boldsymbol{\nu},\boldsymbol{\alpha}\rangle =m
\end{subarray}}
\frac{1-t^{1+n-\langle \boldsymbol{\rho},\boldsymbol{\alpha}
\rangle}}{1-t^{n-\langle \boldsymbol{\rho},\boldsymbol{\alpha}
\rangle}},
\end{equation}
respectively, which---upon inserting in Eq. \eqref{ps}---gives rise
to Eq. \eqref{coefV}.
\end{proof}

\begin{remark}
The stabilizer subgroups
$\hat{\mathcal{S}}_{n,\boldsymbol{\lambda}}^{(m)}$ and
$\hat{\mathcal{S}}_{n,\boldsymbol{\lambda}}^{(m)}\cap
\hat{\mathcal{S}}_{n,\boldsymbol{\lambda}+\boldsymbol{\nu}}^{(m)}$
in the proof of Proposition \ref{action:prp} consist of (direct
products of) permutation groups. It is well-known (and readily seen
by induction) that the Poincar\'e series of the permutation group
$\mathcal{S}_{\ell}$ admits the product representation
$\prod_{j=1}^\ell (1-t^j)/(1-t)$  (cf. e.g. Ref.
\cite{mac:symmetric}, Chapter III, \S 1). With the aid of this
latter product formula it is not so difficult to verify Eqs.
\eqref{ps1}, \eqref{ps2} (and thus Eq. \eqref{coefV}) directly (i.e.
without invoking the much more general results of Macdonald
\cite{mac:poincare} cited in the proof). (We thank the referee for
making this point.)
\end{remark}

\subsection{Hilbert Space Structure}
We shall now endow the function space
$\mathcal{C}(\mathcal{P}^{(m)})$ with a inner product, turning it
into a (finite-dimensional) Hilbert space
$\mathcal{H}^{(m)}:=\ell^2(\mathcal{P}^{(m)},\Delta^{(m)})$
characterized by a positive weight function
$\Delta^{(m)}:\mathcal{P}^{(m)}\to (0,\infty)$. To this end it will
always be assumed from here onwards that the coupling parameter $t$
lies in the repulsive regime
\begin{equation}\label{rep}
-1<t<1
\end{equation}
(unless explicitly stated otherwise). For two arbitrary functions
$\psi,\phi\in\mathcal{C}(\mathcal{P}^{(m)})$ the inner product in
question is then defined as
\begin{subequations}
\begin{equation}
\langle \psi,\phi\rangle^{(m)}:=
\sum_{\boldsymbol{\lambda}\in\mathcal{P}^{(m)}}
\psi_{\boldsymbol{\lambda}}\overline{\phi_{\boldsymbol{\lambda}}}\;
\Delta_{\boldsymbol{\lambda}}^{(m)} ,
\end{equation}
where the weight function is given by
\begin{equation}
\Delta^{(m)}_{\boldsymbol{\lambda}} :=
\prod_{\begin{subarray}{c}\boldsymbol{\alpha}\in \mathbf{R}^+\\
\langle \boldsymbol{\lambda},\boldsymbol{\alpha}\rangle =0
\end{subarray}}
\frac{1-t^{\langle \boldsymbol{\rho},\boldsymbol{\alpha}
\rangle}}{1-t^{1+\langle \boldsymbol{\rho},\boldsymbol{\alpha}
\rangle}}
\prod_{\begin{subarray}{c}\boldsymbol{\alpha}\in \mathbf{R}^+\\
\langle \boldsymbol{\lambda},\boldsymbol{\alpha}\rangle =m
\end{subarray}}
\frac{1-t^{n-\langle \boldsymbol{\rho},\boldsymbol{\alpha}
\rangle}}{1-t^{1+n-\langle \boldsymbol{\rho},\boldsymbol{\alpha}
\rangle}}  .
\end{equation}
\end{subequations}
(Notice in this connection that the restriction of the coupling
parameter to the repulsive regime \eqref{rep} ensures that the
values of the weight function $\Delta^{(m)}_{\boldsymbol{\lambda}}$
are indeed positive for all
$\boldsymbol{\lambda}\in\mathcal{P}^{(m)}$.)

\begin{proposition}[Adjoint]\label{adjoint:prp}
The Laplace operators $ L^{(m)}_k$ and $ L^{(m)}_{n-k}$ ($k\in\{
1,\ldots ,n-1\}$) are each others adjoints in $\mathcal{H}^{(m)}$
\begin{equation}
\forall \psi,\phi\in\mathcal{C}(\mathcal{P}^{(m)}):\qquad \langle
L^{(m)}_k\psi ,\phi\rangle^{(m)} = \langle \psi
,L^{(m)}_{n-k}\phi\rangle^{(m)} .
\end{equation}
\end{proposition}
\begin{proof}
The proof hinges on the explicit formula for the action of
$L^{(m)}_k$ in Proposition \ref{action:prp}.
 Elementary manipulations reveal that
\begin{eqnarray*} \langle
L^{(m)}_k\psi,\phi\rangle^{(m)} &=&
\sum_{\boldsymbol{\lambda}\in\mathcal{P}^{(m)}}
(L^{(m)}_k\psi)_{\boldsymbol{\lambda}}
\overline{\phi_{\boldsymbol{\lambda}}}\Delta^{(m)}_{\boldsymbol{\lambda}}
\\
&=&\sum_{\boldsymbol{\nu}\in\mathcal{S}_n(\omega_k)}
\sum_{\begin{subarray}{c}\boldsymbol{\lambda}\in\mathcal{P}^{(m)} \\
\boldsymbol{\lambda}+\boldsymbol{\nu}\in\mathcal{P}^{(m)}
\end{subarray}}
V^{(m)}_{\boldsymbol{\lambda},
\boldsymbol{\nu}}\psi_{\boldsymbol{\lambda}+\boldsymbol{\nu}}
\overline{\phi_{\boldsymbol{\lambda}}}\Delta^{(m)}_{\boldsymbol{\lambda}}
\\
&\stackrel{(i)}{=}&\sum_{\boldsymbol{\nu}\in\mathcal{S}_n(\omega_k)}
\sum_{\begin{subarray}{c}\boldsymbol{\mu}\in\mathcal{P}^{(m)} \\
\boldsymbol{\mu}-\boldsymbol{\nu}\in\mathcal{P}^{(m)}
\end{subarray}}
\psi_{\boldsymbol{\mu}}
V^{(m)}_{\boldsymbol{\mu}-\boldsymbol{\nu},\boldsymbol{\nu}}
\overline{\phi_{\boldsymbol{\mu}-\boldsymbol{\nu}}}
\Delta^{(m)}_{\boldsymbol{\mu}-\boldsymbol{\nu}}
\\
&\stackrel{(ii)}{=}&\sum_{\boldsymbol{\nu}\in\mathcal{S}_n(\omega_k)}
\sum_{\begin{subarray}{c}\boldsymbol{\mu}\in\mathcal{P}^{(m)} \\
\boldsymbol{\mu}-\boldsymbol{\nu}\in\mathcal{P}^{(m)}
\end{subarray}}
\psi_{\boldsymbol{\mu}} V^{(m)}_{\boldsymbol{\mu},-\boldsymbol{\nu}}
\overline{\phi_{\boldsymbol{\mu}-\boldsymbol{\nu}}}\Delta^{(m)}_{\boldsymbol{\mu}}
\\
&\stackrel{(iii)}{=}& \sum_{\boldsymbol{\mu}\in\mathcal{P}^{(m)}}
\psi_{\boldsymbol{\mu}}
\overline{(L^{(m)}_{n-k}\phi)_{\boldsymbol{\mu}}}\Delta^{(m)}_{\boldsymbol{\mu}}
= \langle \psi,L^{(m)}_{n-k}\phi\rangle^{(m)} ,
\end{eqnarray*}
where we have used: {\em (i)} the substitution
$\boldsymbol{\lambda}=\boldsymbol{\mu}-\boldsymbol{\nu}$, {\em (ii)}
the identity
\begin{equation}\label{pears}
V^{(m)}_{\boldsymbol{\mu}-\boldsymbol{\nu},\boldsymbol{\nu}}\,
\Delta^{(m)}_{\boldsymbol{\mu}-\boldsymbol{\nu}}
= V^{(m)}_{\boldsymbol{\mu},-\boldsymbol{\nu}}\,
\Delta^{(m)}_{\boldsymbol{\mu}} ,
\end{equation}
and {\em (iii)} the facts that the coefficient
$V^{(m)}_{\boldsymbol{\mu},-\boldsymbol{\nu}}$ is real and
$-\boldsymbol{\omega}_k\in\mathcal{S}_n(\boldsymbol{\omega}_{n-k})$.
To infer the identity in Eq. \eqref{pears} one observes that for
$\boldsymbol{\mu}\in\mathcal{P}^{(m)}$ and
$\boldsymbol{\nu}\in\mathcal{S}_n(\boldsymbol{\omega}_k)$ such that
$\boldsymbol{\mu}-\boldsymbol{\nu}\in\mathcal{P}^{(m)}$ both sides
$V^{(m)}_{\boldsymbol{\mu}-\boldsymbol{\nu},\boldsymbol{\nu}}\,
\Delta^{(m)}_{\boldsymbol{\mu}-\boldsymbol{\nu}}$
and $V^{(m)}_{\boldsymbol{\mu},-\boldsymbol{\nu}}\,
\Delta^{(m)}_{\boldsymbol{\mu}}$ reduce---upon canceling common
terms from the numerator and denominator---to
\begin{equation*}
\prod_{\begin{subarray}{c}\boldsymbol{\alpha}\in \mathbf{R}^+\\
\langle \boldsymbol{\mu},\boldsymbol{\alpha}\rangle =0\\
\langle \boldsymbol{\nu},\boldsymbol{\alpha}\rangle =0
\end{subarray}}
\frac{1-t^{\langle \boldsymbol{\rho},\boldsymbol{\alpha}
\rangle}}{1-t^{1+\langle \boldsymbol{\rho},\boldsymbol{\alpha}
\rangle}}
\prod_{\begin{subarray}{c}\boldsymbol{\alpha}\in \mathbf{R}^+\\
\langle \boldsymbol{\mu},\boldsymbol{\alpha}\rangle =m\\
\langle \boldsymbol{\nu},\boldsymbol{\alpha}\rangle =0
\end{subarray}}
\frac{1-t^{n-\langle \boldsymbol{\rho},\boldsymbol{\alpha}
\rangle}}{1-t^{1+n-\langle \boldsymbol{\rho},\boldsymbol{\alpha}
\rangle}} .
\end{equation*}
\end{proof}

\section{Bethe Ansatz Eigenfunctions}\label{sec3}
In this section the eigenfunctions of the Laplacian $L^{(m)}_k$
\eqref{lapk}, \eqref{bconv} are constructed by means of the Bethe
Ansatz method of Lieb and Liniger
\cite{lie-lin:exact,mat:many-body,gau:fonction,kor-bog-ize:quantum}.

\subsection{Bethe Ansatz}
If we ignore boundary effects for a moment and interpret $L^{(m)}_k$
\eqref{lapk} ({\em without} the boundary convention \eqref{bconv})
as a Laplacian acting on functions $\psi :\mathcal{P}\to\mathbb{C}$,
then clearly the plane wave
$\psi_{\boldsymbol{\lambda}}(\boldsymbol{\xi})=\exp (i\langle
\boldsymbol{\lambda},\boldsymbol{\xi}\rangle )$ with wave number
$\boldsymbol{\xi}\in\boldsymbol{E}/(2\pi \mathcal{Q})$ constitutes
an eigenfunction corresponding to the eigenvalue
$E_k(\boldsymbol{\xi})=
\sum_{\boldsymbol{\nu}\in\mathcal{S}_n(\boldsymbol{\omega}_k)}
\exp (i\langle \boldsymbol{\nu},\boldsymbol{\xi}\rangle )$. The
Bethe Ansatz method aims to construct the eigenfunctions $\psi
:\mathcal{P}^{(m)}\to\mathbb{C}$ for the operator $L^{(m)}_k$
\eqref{lapk} {\em with} the boundary convention \eqref{bconv} via a
suitable linear combinations of plane waves (corresponding to the
same eigenvalue $E_k(\boldsymbol{\xi})$). This prompts us to look
for eigenfunctions given by a linear combination of plane waves
$\exp (i\langle \boldsymbol{\lambda},\boldsymbol{\xi}_\sigma\rangle
)$, $\sigma\in\mathcal{S}_n$ with coefficients such that the
boundary conditions in Proposition \ref{bref:prp} are satisfied.
Specifically, we will employ a $\mathcal{S}_n$-invariant (in
$\boldsymbol{\xi}$) Bethe Ansatz wave function of the form
\begin{subequations}
\begin{equation}\label{bawf}
\Psi_{\boldsymbol{\lambda}} (\boldsymbol{\xi}) = \frac{1}{\delta
(\boldsymbol{\xi})} \sum_{\sigma\in \mathcal{S}_n} (-1)^{\sigma}
\mathcal{C}(\boldsymbol{\xi}_\sigma) e^{i\langle \boldsymbol{\rho}
+\boldsymbol{\lambda} ,\boldsymbol{\xi}_\sigma \rangle}, \qquad
\boldsymbol{\xi}\in 2\pi \text{Int}(\boldsymbol{A}),
\end{equation}
where $\boldsymbol{\xi}_\sigma:=\sigma (\boldsymbol{\xi})$,
$(-1)^\sigma:=\det(\sigma)=(-1)^{\ell (\sigma)}$,
\begin{equation}\label{delta}
\delta (\boldsymbol{\xi}) := \prod_{\boldsymbol{\alpha}\in
\mathbf{R}^+} (e^{i\langle
\boldsymbol{\alpha},\boldsymbol{\xi}\rangle/2}-e^{-i\langle
\boldsymbol{\alpha},\boldsymbol{\xi}\rangle/2} ) ,
\end{equation}
\end{subequations}
and $\text{Int}(\boldsymbol{A})=\{
\boldsymbol{\xi}\in\boldsymbol{E}\mid \langle
\boldsymbol{\xi},\boldsymbol{\alpha}_0\rangle < 1 ,\; \langle
\boldsymbol{\xi},\boldsymbol{\alpha}_j\rangle > 0,\, j=1,\ldots ,n-1
\}$. (The condition that $\boldsymbol{\xi}\in 2\pi
\text{Int}(\boldsymbol{A})$ guarantees that the denominator $\delta
(\boldsymbol{\xi})$ is nonzero.)

\begin{proposition}[Bethe Wave Function]\label{bwf:prp}
The Bethe Ansatz wave function $\Psi_\lambda (\boldsymbol{\xi})$
\eqref{bawf}, \eqref{delta} satisfies the boundary reflection
relations in Proposition \ref{bref:prp} for $j=1,\ldots ,n-1$
provided that
\begin{equation}\label{Cf}
\mathcal{C}(\boldsymbol{\xi}) = \prod_{\boldsymbol{\alpha}\in
\mathbf{R}^+} (1-t\, e^{-i\langle \boldsymbol{\alpha}
,\boldsymbol{\xi}\rangle})
\end{equation}
(or a scalar multiple thereof).
\end{proposition}

\begin{proof}
Let $\boldsymbol{\lambda}\in\mathcal{P}^{(m)}$ and
$\boldsymbol{\nu}\in\mathcal{S}_n(\boldsymbol{\omega}_k)$ such that
$\langle
\boldsymbol{\lambda}+\boldsymbol{\nu},\boldsymbol{\alpha}_j\rangle
=-1$ for some $j\in \{ 1,\ldots ,n-1\}$. Equating
\begin{equation*}
\Psi_{\boldsymbol{\lambda}+\boldsymbol{\nu}} (\boldsymbol{\xi}) =
\frac{1}{\delta (\boldsymbol{\xi})} \sum_{\sigma\in \mathcal{S}_n}
(-1)^{\sigma} \mathcal{C}(\boldsymbol{\xi}_\sigma ) e^{i\langle
\boldsymbol{\rho} +\boldsymbol{\lambda}+\boldsymbol{\nu}
,\boldsymbol{\xi}_\sigma \rangle}
\end{equation*}
to
\begin{eqnarray*}
t\, \Psi_{r_j(\boldsymbol{\lambda}+\boldsymbol{\nu})}
(\boldsymbol{\xi}) &=&
t\, \Psi_{\boldsymbol{\lambda}+\boldsymbol{\nu}+
\boldsymbol{\alpha}_j} (\boldsymbol{\xi}) \\
&=& \frac{t}{\delta (\boldsymbol{\xi})} \sum_{\sigma\in
\mathcal{S}_n} (-1)^{\sigma}
\mathcal{C}(\boldsymbol{\xi}_\sigma)e^{i\langle
\boldsymbol{\alpha}_j ,\boldsymbol{\xi}_\sigma \rangle} e^{i\langle
\boldsymbol{\rho} +\boldsymbol{\lambda}+\boldsymbol{\nu}
,\boldsymbol{\xi}_\sigma \rangle}
\end{eqnarray*}
leads to the relation
\begin{equation*}
\sum_{\sigma\in
\mathcal{S}_{n,\boldsymbol{\rho}+\boldsymbol{\lambda}+\boldsymbol{\nu}}}
(-1)^{\sigma} \mathcal{C}(\boldsymbol{\xi}_{\sigma\tau})= t
\sum_{\sigma\in
\mathcal{S}_{n,\boldsymbol{\rho}+\boldsymbol{\lambda}+\boldsymbol{\nu}}}
(-1)^{\sigma} \mathcal{C}(\boldsymbol{\xi}_{\sigma\tau})e^{i\langle
\boldsymbol{\alpha}_j ,\boldsymbol{\xi}_{\sigma\tau}\rangle} \qquad
\forall \tau\in\mathcal{S}_n.
\end{equation*}
Because $r_j$ stabilizes
$\boldsymbol{\rho}+\boldsymbol{\lambda}+\boldsymbol{\nu}$ (i.e.
$r_j\in
\mathcal{S}_{n,\boldsymbol{\rho}+\boldsymbol{\lambda}+\boldsymbol{\nu}}$),
the latter relation can be rewritten as
\begin{eqnarray*}
\lefteqn{\sum_{\begin{subarray}{c}
\sigma\in \mathcal{S}_{n,\boldsymbol{\rho}+\boldsymbol{\lambda}+\boldsymbol{\nu}}\\
\sigma^{-1}(\boldsymbol{\alpha}_j)\in\mathbf{R}^+\end{subarray}}
(-1)^{\sigma}
[\mathcal{C}(\boldsymbol{\xi}_{\sigma\tau})-
\mathcal{C}(r_j(\boldsymbol{\xi}_{\sigma\tau}))]}
&& \\
&& = t\,\sum_{\begin{subarray}{c}
\sigma\in \mathcal{S}_{\boldsymbol{\rho}+\boldsymbol{\lambda}+\boldsymbol{\nu}}\\
\sigma^{-1}(\boldsymbol{\alpha}_j)\in\mathbf{R}^+\end{subarray}}
(-1)^{\sigma} [
\mathcal{C}(\boldsymbol{\xi}_{\sigma\tau})e^{i\langle
\boldsymbol{\alpha}_j
,\boldsymbol{\xi}_{\sigma\tau}\rangle}-
\mathcal{C}(r_j(\boldsymbol{\xi}_{\sigma\tau}))e^{-i\langle
\boldsymbol{\alpha}_j ,\boldsymbol{\xi}_{\sigma\tau}\rangle}] .
\end{eqnarray*}
By varying $\boldsymbol{\lambda}$ and $\boldsymbol{\nu}$ it is seen
that this relation implies that
\begin{equation*}
\mathcal{C}(\boldsymbol{\xi})-\mathcal{C}(r_j(\boldsymbol{\xi}))=
t\, [ \mathcal{C}(\boldsymbol{\xi})e^{i\langle \boldsymbol{\alpha}_j
,\boldsymbol{\xi}\rangle}-\mathcal{C}(r_j(\boldsymbol{\xi}))e^{-i\langle
\boldsymbol{\alpha}_j ,\boldsymbol{\xi}\rangle}]
\end{equation*}
(as an identity in $\boldsymbol{\xi}$) for all reflections $r_j$,
$j=1,\ldots, n-1$, or equivalenty (assuming
$\mathcal{C}(\boldsymbol{\xi})$ is nontrivial in the sense that it
does not vanish identically)
\begin{equation*}
\frac{\mathcal{C}(\boldsymbol{\xi})}{\mathcal{C}(r_j(\boldsymbol{\xi}))}
=\frac{1-t e^{-i\langle \boldsymbol{\alpha}_j
,\boldsymbol{\xi}\rangle}}{1-t e^{i\langle \boldsymbol{\alpha}_j
,\boldsymbol{\xi}\rangle}},\qquad j=1,\ldots ,n-1.
\end{equation*}
Hence $\mathcal{C}(\boldsymbol{\xi})$ must be of the form
\begin{equation*}
\mathcal{C}(\boldsymbol{\xi}) =
c_0(\boldsymbol{\xi})\prod_{\boldsymbol{\alpha}\in \mathbf{R}^+}
(1-t e^{-i\langle \boldsymbol{\alpha} ,\boldsymbol{\xi}\rangle}),
\end{equation*}
where $c_0(\boldsymbol{\xi})$ denotes an arbitrary
$\mathcal{S}_n$-invariant overall factor (i.e.
$c_0(\boldsymbol{\xi}_\sigma)=c_0(\boldsymbol{\xi})$, $\forall
\sigma\in \mathcal{S}_n$).
\end{proof}

\subsection{Bethe Equations}
By pulling the overall factor $\delta (\boldsymbol{\xi})$ inside the
sum and exploiting the anti-invariance $\delta
(\boldsymbol{\xi}_\sigma)=(-1)^\sigma \delta (\boldsymbol{\xi})$,
the Bethe wave function $\Psi_{\boldsymbol{\lambda}}
(\boldsymbol{\xi}) $ \eqref{bawf}, \eqref{delta}, with coefficients
$\mathcal{C}(\boldsymbol{\xi})$ taken from Eq. \eqref{Cf}, passes
over to
\begin{equation}\label{hwf}
\Psi_{\boldsymbol{\lambda}} (\boldsymbol{\xi}) =  \sum_{\sigma\in
\mathcal{S}_n}\Bigl( \prod_{\boldsymbol{\alpha}\in \mathbf{R}^+}
\frac{1-t\, e^{-i\langle
\boldsymbol{\alpha},\boldsymbol{\xi}_\sigma\rangle} }{1-e^{-i\langle
\boldsymbol{\alpha},\boldsymbol{\xi}_\sigma\rangle} }\Bigr)
e^{i\langle \boldsymbol{\lambda} ,\boldsymbol{\xi}_\sigma \rangle} .
\end{equation}
From this expression it is clear that---for $\boldsymbol{\lambda}$
fixed---the Bethe wave function amounts to a Hall-Littlewood
polynomial in the spectral parameter $\boldsymbol{\xi}$
\cite{mac:symmetric,mac:orthogonal}. We will now derive conditions
on the spectral parameter such that the Bethe wave function
satisfies the boundary reflection relations in Eqs. \eqref{bref1},
\eqref{bref2} for $j=0$.

\begin{proposition}[Bethe System]\label{bs:prp}
The Bethe Ansatz wave function $\Psi_\lambda (\boldsymbol{\xi})$
\eqref{hwf} satisfies the boundary reflection relations in
Proposition \ref{bref:prp} for $j=0$ if the spectral parameter
$\boldsymbol{\xi}\in 2\pi\text{Int}(\boldsymbol{A})$ solves the
algebraic system
\begin{equation}\label{bs}
e^{im\langle \boldsymbol{\beta},\boldsymbol{\xi}   \rangle } =
\Bigl( \frac{1-t\,
e^{i\langle\boldsymbol{\beta},\boldsymbol{\xi}\rangle}} {
e^{i\langle\boldsymbol{\beta},\boldsymbol{\xi}\rangle}-t}\Bigr)^2
\prod_{\begin{subarray}{c}\boldsymbol{\alpha}\in \mathbf{R}\\
\langle \boldsymbol{\alpha},\boldsymbol{\beta}\rangle =1
\end{subarray}}
\frac{1-t\, e^{i\langle\boldsymbol{\alpha},\boldsymbol{\xi}\rangle}}
{e^{i\langle\boldsymbol{\alpha},\boldsymbol{\xi}\rangle}-t}, \qquad
\forall \boldsymbol{\beta}\in\mathbf{R}.
\end{equation}
\end{proposition}

\begin{proof}
Let us write
$\hat{\mathcal{C}}(\boldsymbol{\xi}):=\prod_{\boldsymbol{\alpha}\in
\mathbf{R}^+} \frac{1-t\, e^{-i\langle
\boldsymbol{\alpha},\boldsymbol{\xi}\rangle} }{1-e^{-i\langle
\boldsymbol{\alpha},\boldsymbol{\xi}\rangle} }$ and let
$\boldsymbol{\lambda}\in\mathcal{P}^{(m)}$ and
$\boldsymbol{\nu}\in\mathcal{S}_n(\boldsymbol{\omega}_k)$ such that
$\langle
\boldsymbol{\lambda}+\boldsymbol{\nu},\boldsymbol{\alpha}_0\rangle
=m+1$. Equating
\begin{equation*}
\Psi_{\boldsymbol{\lambda}+\boldsymbol{\nu}} (\boldsymbol{\xi}) =
\sum_{\sigma\in
\mathcal{S}_n}\hat{\mathcal{C}}(\boldsymbol{\xi}_\sigma) e^{i\langle
\boldsymbol{\lambda} +\boldsymbol{\nu},\boldsymbol{\xi}_\sigma
\rangle}
\end{equation*}
to
\begin{equation*}
t\, \Psi_{r_0^{(m)}(\boldsymbol{\lambda}+\boldsymbol{\nu})}
(\boldsymbol{\xi}) = t \sum_{\sigma\in
\mathcal{S}_n}\hat{\mathcal{C}}(\boldsymbol{\xi}_\sigma) e^{i\langle
r_0^{(m)}(\boldsymbol{\lambda}
+\boldsymbol{\nu}),\boldsymbol{\xi}_\sigma \rangle}
\end{equation*}
yields the relation
\begin{equation*}
\sum_{\sigma\in \mathcal{S}_n}\bigl( 1-t\, e^{-i\langle
\boldsymbol{\alpha}_0,\boldsymbol{\xi}_\sigma
\rangle}\bigr)\hat{\mathcal{C}}(\boldsymbol{\xi}_\sigma) e^{i\langle
\boldsymbol{\lambda} +\boldsymbol{\nu},\boldsymbol{\xi}_\sigma
\rangle} =0 ,
\end{equation*}
or equivalently
\begin{eqnarray*}
&&\sum_{\begin{subarray}{c} \sigma\in \mathcal{S}_n\\
\sigma^{-1}(\boldsymbol{\alpha}_0)\in\mathbf{R}^+\end{subarray} }
\Bigl( \bigl( 1-t\, e^{-i\langle
\boldsymbol{\alpha}_0,\boldsymbol{\xi}_\sigma
\rangle}\bigr)\hat{\mathcal{C}}(\boldsymbol{\xi}_\sigma) e^{i\langle
\boldsymbol{\lambda} +\boldsymbol{\nu},\boldsymbol{\xi}_\sigma
\rangle} \; + \\
&&  \makebox[5em]{}  \bigl( 1-t\, e^{-i\langle
\boldsymbol{\alpha}_0,r_0^{(0)}(\boldsymbol{\xi}_\sigma)
\rangle}\bigr)\hat{\mathcal{C}}(r_0^{(0)}(\boldsymbol{\xi}_\sigma))
e^{i\langle \boldsymbol{\lambda}
+\boldsymbol{\nu},r_0^{(0)}(\boldsymbol{\xi}_\sigma)\rangle}\Bigr)
=0 ,
\end{eqnarray*}
where $r_0^{(0)}\in\mathcal{S}_n$ denotes the orthogonal reflection
$r_0^{(0)}(\mathbf{x})=\mathbf{x}-\langle
\mathbf{x},\boldsymbol{\alpha}_0\rangle\boldsymbol{\alpha}_0$. The
latter equation translates to
\begin{eqnarray*}
\lefteqn{\sum_{\begin{subarray}{c} \sigma\in \mathcal{S}_n\\
\sigma^{-1}(\boldsymbol{\alpha}_0)\in\mathbf{R}^+\end{subarray} }
\Bigl( \bigl( 1-t\, e^{-i\langle
\boldsymbol{\alpha}_0,\boldsymbol{\xi}_\sigma
\rangle}\bigr)\hat{\mathcal{C}}(\boldsymbol{\xi}_\sigma) \; +} && \\
&&   \bigl( 1-t\, e^{i\langle
\boldsymbol{\alpha}_0,\boldsymbol{\xi}_\sigma
\rangle}\bigr)\hat{\mathcal{C}}(r_0^{(0)}(\boldsymbol{\xi}_\sigma))
e^{-(m+1)i\langle \boldsymbol{\alpha}_0,\boldsymbol{\xi}_\sigma
\rangle} \Bigr) e^{i\langle \boldsymbol{\lambda}
+\boldsymbol{\nu},\boldsymbol{\xi}_\sigma\rangle} =0 ,
\end{eqnarray*}
which is satisfied if
\begin{equation*}
\bigl( 1-t\, e^{-i\langle
\boldsymbol{\alpha}_0,\boldsymbol{\xi}_\sigma
\rangle}\bigr)\hat{\mathcal{C}}(\boldsymbol{\xi}_\sigma) + \bigl(
1-t\, e^{i\langle \boldsymbol{\alpha}_0,\boldsymbol{\xi}_\sigma
\rangle}\bigr)\hat{\mathcal{C}}(r_0^{(0)}(\boldsymbol{\xi}_\sigma))
e^{-(m+1)i\langle \boldsymbol{\alpha}_0,\boldsymbol{\xi}_\sigma
\rangle} =0
\end{equation*}
for all $\sigma\in\mathcal{S}_n$, or equivalently (assuming
$\hat{\mathcal{C}}(\boldsymbol{\xi}_\sigma)\neq 0$)
\begin{equation*}
e^{(m+1)i\langle \boldsymbol{\beta},\boldsymbol{\xi} \rangle}
=-\frac{\hat{\mathcal{C}}(r_0^{(0)}(\boldsymbol{\xi}_\sigma))}
{\hat{\mathcal{C}}(\boldsymbol{\xi}_\sigma)}
 \frac{1-t\, e^{i\langle
\boldsymbol{\beta},\boldsymbol{\xi}\rangle} }{1-t\, e^{-i\langle
\boldsymbol{\beta},\boldsymbol{\xi} \rangle}},\qquad \forall
\sigma\in\mathcal{S}_n,
\end{equation*}
where $\boldsymbol{\beta}:=\sigma^{-1} (\boldsymbol{\alpha}_0)$. The
proposition now follows upon inserting
\begin{equation*}
\frac{\hat{\mathcal{C}}(r_0^{(0)}(\boldsymbol{\xi}_\sigma))}
{\hat{\mathcal{C}}(\boldsymbol{\xi}_\sigma)}
=-
\prod_{\begin{subarray}{c}\boldsymbol{\alpha}\in \mathbf{R}^+\\
\langle \boldsymbol{\alpha},\boldsymbol{\alpha}_0\rangle >0
\end{subarray}}
\frac{1-t\,
e^{i\langle\boldsymbol{\alpha},\boldsymbol{\xi}_\sigma\rangle}} {
e^{i\langle\boldsymbol{\alpha},\boldsymbol{\xi}_\sigma \rangle}-t}
=-
\prod_{\begin{subarray}{c}\boldsymbol{\alpha}\in \mathbf{R}\\
\langle \boldsymbol{\alpha},\boldsymbol{\beta}\rangle >0
\end{subarray}}
\frac{1-t\, e^{i\langle\boldsymbol{\alpha},\boldsymbol{\xi}\rangle}}
{e^{i\langle\boldsymbol{\alpha},\boldsymbol{\xi}\rangle}-t} .
\end{equation*}
\end{proof}

\section{Solution of the Bethe Equations}\label{sec4}
In this section the Bethe System in Proposition \ref{bs:prp} is
solved using a variational technique due to C.N. Yang and C.P. Yang
\cite{yan-yan:thermodynamics,mat:many-body,gau:fonction,kor-bog-ize:quantum}.

\subsection{Solution}
The following theorem provides (the existence of) a sequence
$\boldsymbol{\xi}_{\boldsymbol{\mu}}\in
2\pi\text{Int}(\boldsymbol{A})$ of solutions to the Bethe system in
Proposition \ref{bs:prp} labeled by vectors (playing the role of
quantum numbers) $\boldsymbol{\mu}\in\mathcal{P}^{(m)}$.

\begin{theorem}[Bethe Vectors]\label{bethe-sol:thm}
For each $\boldsymbol{\mu}\in \mathcal{P}^{(m)}$ there exists a
(unique) Bethe vector $\boldsymbol{\xi}_{\boldsymbol{\mu}}\in 2\pi
\text{Int}(\boldsymbol{A})=\{ \mathbf{x}\in\boldsymbol{E}\mid
\langle \mathbf{x},\boldsymbol{\alpha}_0\rangle < 2\pi ,\; \langle
\mathbf{x},\boldsymbol{\alpha}_j\rangle > 0,\, j=1,\ldots ,n-1 \}$
such that $\boldsymbol{\xi}_{\boldsymbol{\mu}}$ satisfies the system
in Eq. \eqref{bs}. Moreover, these Bethe vectors have the following
properties:
\begin{itemize}
\item[{\em (i)}]
$\boldsymbol{\xi}_{\boldsymbol{\mu}^\prime}=
\boldsymbol{\xi}_{\boldsymbol{\mu}}$
if and only if $\boldsymbol{\mu}^\prime=\boldsymbol{\mu}$,
\item[{\em (ii)}] $\boldsymbol{\xi}_{\boldsymbol{\mu}}$ depends
smoothly on the boundary parameter $t\in(-1,1)$,
\item[{\em (iii)}]
$\boldsymbol{\xi}_{\boldsymbol{\mu}}=
\frac{2\pi}{n+m}(\boldsymbol{\rho}+\boldsymbol{\mu})$
for $t=0$.
\end{itemize}
\end{theorem}

\subsection{Proof}
In standard coordinates the Bethe system of Proposition \ref{bs:prp}
reads
\begin{equation}\label{beqc}
e^{im(\xi_j-\xi_k)}= \prod_{\begin{subarray}{c} 1\leq \ell\leq n
\\ \ell \neq j\end{subarray}}
\frac{1-te^{i(\xi_j-\xi_\ell)}}{e^{i(\xi_j-\xi_\ell)}-t}
\prod_{\begin{subarray}{c} 1\leq \ell\leq n
\\ \ell \neq k\end{subarray}}
\frac{1-te^{i(\xi_\ell-\xi_k)}}{e^{i(\xi_\ell-\xi_k)}-t},
\end{equation}
for $1\leq j\neq k\leq n$. This overdetermined system of $n(n-1)$
equations in the variables $\xi_1,\ldots ,\xi_n$ is equivalent to
the system of $n$ equations
\begin{equation}\label{beqp}
e^{im\xi_j}= c \prod_{\begin{subarray}{c} 1\leq \ell\leq n
\\ \ell \neq j\end{subarray}}
\frac{1-te^{i(\xi_j-\xi_\ell)}}{e^{i(\xi_j-\xi_\ell)}-t}, \qquad
j=1,\ldots ,n,
\end{equation}
where $c\neq 0$ denotes an overall constant factor that we can scale
to $1$ by means of the translation $\xi_j\to\xi_j-im^{-1}\log c$,
$j=1,\ldots n$. Picking thus $c=1$ and taking the logarithm of both
sides recasts Eq. \eqref{beqp} in the additive form
\begin{equation}\label{beq}
m\xi_j+\sum_{\begin{subarray}{c} 1\leq \ell\leq n
\\ \ell \neq j\end{subarray}} \theta (\xi_j-\xi_\ell) = 2\pi m_j,
\qquad j=1,\ldots ,n,
\end{equation}
where $\mathbf{m}=(m_1,\ldots ,m_n)\in \mathbb{Z}^n$ and
\begin{subequations}
\begin{eqnarray}\label{theta1}
\theta (x) &:= & (1-t^2)\int_0^x (1-2t\cos (x)+t^2)^{-1} \text{d}x
\\
&=& 2\arctan \left( \frac{1+t}{1-t}\tan\bigl(\frac{x}{2}\bigr) \right)
 \label{theta2}\\
&=& i\log \left( \frac{1-te^{ix}}{e^{ix}-t} \right) .\label{theta3}
\end{eqnarray}
\end{subequations}
Here the branches of the arctangent function and those of the
logarithmic function are to be chosen in such a way that {\em (i)}
$\theta (x)$ \eqref{theta2}, \eqref{theta3} is quasi-periodic:
$\theta (x+2\pi)=\theta (x)+2\pi$, and {\em (ii)} $\theta (x)$
\eqref{theta2}, \eqref{theta3} varies from $-\pi$ to $\pi$ as $x$
varies from $-\pi$ to $\pi$ (which corresponds to the principal
branch). We notice that this choice of the branches ensures that
$\theta (x)$ \eqref{theta2}, \eqref{theta3} is smooth on the whole
real axis and strictly monotonously increasing.

\begin{lemma}\label{sol:lem}
For each $n$-tuple $\mathbf{m}=(m_1,\ldots ,m_n)\in\mathbb{Z}^n$,
there exists a unique vector
$\boldsymbol{\xi}(\mathbf{m})=(\xi_1(\mathbf{m}),\ldots
,\xi_n(\mathbf{m}))$ solving the system in Eq. \eqref{beq} (with
$\theta (x)$ of the form in Eqs.  \eqref{theta1}--\eqref{theta3}).
Furthermore, this solution $\boldsymbol{\xi} (\mathbf{m})$ depends
smoothly on the boundary parameter $t\in (-1,1)$.
\end{lemma}

\begin{proof}
Let
\begin{equation}\label{V}
V(\xi_1,\ldots ,\xi_n):= \frac{m}{2}\sum_{j=1}^n \xi_j^2 +
\frac{1}{2} \sum_{j,k =1}^n \Theta (\xi_j-\xi_k )  -2\pi
\sum_{j=1}^n m_j\xi_j ,
\end{equation}
where $\Theta (x) :=\int_0^x \theta (x) \text{d} x$. Clearly the
solution(s) of the system in Eq. \eqref{beq} coincide with the
critical point(s) of the (smooth) function $V(\xi_1,\ldots ,\xi_n)$.
The Hesse matrix of $V$ is given by
\begin{equation*}
H_{j,k}=\frac{\partial^2 V}{\partial \xi_j\partial \xi_k}= \left(
m+\sum_{\ell =1}^m \theta^\prime (\xi_j-\xi_\ell) \right)
\delta_{j,k}-\theta^\prime (\xi_j-\xi_k) ,\quad 1\leq j,k\leq n,
\end{equation*}
where $\theta^\prime (x)= (1-t^2) (1-2t\cos (x)+t^2)^{-1}>0$. It is
readily seen that this Hesse matrix is positive definite:
\begin{equation*}
\sum_{j,k=1}^n H_{j,k}\, x_j x_k = m\sum_{j=1}^n x_j^2+
\frac{1}{2}\sum_{j,k=1}^n \theta^\prime (\xi_j-\xi_k) (x_j-x_k)^2
\geq m\sum_{j=1}^n x_j^2 >0
\end{equation*}
(for any nonzero vector $\mathbf{x}\in\mathbb{R}^n$). The function
$V(\xi_1,\ldots ,\xi_n)$ is thus strictly convex, i.e., it admits at
most {\em one} critical point: a global minimum. That such global
minimum $\boldsymbol{\xi}(\mathbf{m})$ indeed exists in our case is
immediate from the observation that $V(\xi_1,\ldots ,\xi_n)\to
+\infty$ when $\|\boldsymbol{\xi}\|\to\infty$. (Notice in this
connection that $\Theta (x)\to +\infty$ for $x\to \pm\infty$.) We
thus conclude that the system in Eq. \eqref{beq} has a unique
solution $\boldsymbol{\xi}(\mathbf{m})$ (given by the global minimum
of $V$). It remains to check that the position of this global
minimum depends smoothly on the boundary parameter $t$. To this end
we notice that the integrand of $\theta (x)$ \eqref{theta1} (which,
incidentally, coincides with the generating function for the
Chebyshev polynomials) is analytic in $t$ for $|t|<1$, and thus so
are the function $V(\xi_1,\ldots ,\xi_n)$ and the system of
equations for the global minimum in Eq. \eqref{beq}. The smoothness
in $\xi_1,\ldots ,\xi_n$ and $t$---combined with the fact that the
Hessian $\det (H_{j,k})$ is positive (i.e. nonvanishing)---now
guarantees that the solution $\boldsymbol{\xi}(\mathbf{m})$ to the
latter system must be smooth in $t\in (-1,1)$ by the implicit
function theorem.
\end{proof}

The following lemma shows that the ordering between the components
of the solution $\xi (\mathbf{m})$ coincides with the ordering of
the components of the labeling vector $\mathbf{m}$.

\begin{lemma}\label{pauli:lem}
Let $\mathbf{m}\in\mathbb{Z}^n$ and let
$\boldsymbol{\xi}(\mathbf{m})$ be the associated solution of Eq.
\eqref{beq} detailed in Lemma \ref{sol:lem}. Then for $m_j\geq
m_{k}$ the following inequalities hold
\begin{subequations}
\begin{equation}\label{paulia}
\frac{2\pi (m_j-m_{k})}{m+n\kappa_-(t)} \leq
\xi_j(\mathbf{m})-\xi_{k}(\mathbf{m}) \leq \frac{2\pi
(m_j-m_{k})}{m+n\kappa_+(t)},
\end{equation}
where $\kappa_\pm (t):=\frac{1-t^2}{(1\pm |t|)^2}>0$. So, one has in
particular that
\begin{equation}\label{paulib}
m_j>m_{k}\Longrightarrow\xi_j(\mathbf{m})>\xi_{k}(\mathbf{m})\qquad
\text{and}\qquad
m_j=m_{k}\Longrightarrow\xi_j(\mathbf{m})=\xi_{k}(\mathbf{m}).
\end{equation}
\end{subequations}
\end{lemma}

\begin{proof}
Let $\mathbf{m}$ in $\mathbb{Z}^n$ with $m_j\geq m_k$. Subtracting
the $k^{th}$ equation from the $j^{th}$ equation of the system in
Eq. \eqref{beq} yields that
\begin{equation}\label{dif}
m(\xi_j-\xi_{k}) + \sum_{\ell =1}^n \bigl( \theta (\xi_j-\xi_\ell)
-\theta (\xi_{k}-\xi_\ell )\bigr) = 2\pi (m_j-m_{k}) .
\end{equation}
Since the r.h.s. of this identity is nonnegative and $\theta (x)$ is
strictly monotonously increasing, it follows that
$\xi_j(\mathbf{m})\geq \xi_k(\mathbf{m})$. Furthermore, from the
formula $\theta (x)-\theta(y)=(1-t^2)\int_y^x
(1-2t\cos(x)+t^2)^{-1}\text{d}x$ (cf. Eq. \eqref{theta1}) it is
immediate that
\begin{equation*}
\kappa_+(t)(x-y)\leq \theta (x)-\theta (y)\leq
\kappa_-(t)(x-y)\qquad \text{for}\quad x\geq y.
\end{equation*}
Application of this upper and lower bound so as to estimate the
terms in the sums on the l.h.s. of Eq. \eqref{dif}, now gives rise
to the inequalities
\begin{equation*}
(m+n\kappa_+(t))(\xi_j(\mathbf{m})-\xi_k(\mathbf{m})) \leq 2\pi
(m_j-m_k)\leq (m+n\kappa_-(t))(\xi_j(\mathbf{m})-\xi_k(\mathbf{m}))
,
\end{equation*}
which completes the proof of Eq. \eqref{paulia} (and thus also that
of Eq. \eqref{paulib}).
\end{proof}

The next lemma improves the upper bound on the distance between the
$\xi_j(\mathbf{m})$  and $\xi_k(\boldsymbol{m})$ stemming from Lemma
\ref{pauli:lem} in the situation that the distance between $m_j$ and
$m_k$ is smaller than $n+m$.

\begin{lemma}\label{dist:lem}
Let $\mathbf{m}\in\mathbb{Z}^n$ such that $m_j-m_k< n+m$ and let
$\boldsymbol{\xi}(\mathbf{m})$ be the associated solution of Eq.
\eqref{beq} detailed in Lemma \ref{sol:lem}. Then
\begin{equation}
\xi_j(\mathbf{m})-\xi_k (\mathbf{m})< 2\pi .
\end{equation}
\end{lemma}

\begin{proof}
Subtracting the $k^{th}$ equation from the $j^{th}$ equation of the
system in Eq. \eqref{beq} leads---upon recalling that $\theta (x)$
is odd---to (cf. Eq. \eqref{dif})
\begin{equation}\label{maxdif}
m(\xi_j-\xi_{k}) + \sum_{\ell =1}^n \bigl( \theta (\xi_j-\xi_\ell)
+\theta (\xi_\ell-\xi_k )\bigr) = 2\pi (m_j-m_{k}) .
\end{equation}
If $\xi_j-\xi_k\geq 2\pi$, then the average of $\xi_j-\xi_\ell$ and
$\xi_\ell-\xi_k$ is $\geq \pi$. Hence $\theta (\xi_j-\xi_\ell)
+\theta (\xi_\ell-\xi_k )\geq 2\pi$, in view of the fact that
$\theta (x)$ is strictly monotonously increasing and $\theta
(\pi+x)+\theta(\pi-x)=2\pi$. Plugging this estimate in Eq.
\eqref{maxdif} reveals that $\xi_j(\mathbf{m})-\xi_k(\mathbf{m})\geq
2\pi$ implies that $2\pi (m_j-m_k)\geq m
(\xi_j(\mathbf{m})-\xi_k(\mathbf{m}))+2\pi n\geq 2\pi (m+n)$, which
completes the proof.
\end{proof}

We will now piece the results of Lemmas
\ref{sol:lem}--\ref{dist:lem} together, so as to arrive at a proof
for Theorem \ref{bethe-sol:thm}.

Let $\mathbf{m}\in\mathbb{Z}^n$ such that
\begin{subequations}
\begin{equation}\label{meq}
m_1>m_2>\cdots >m_n\quad \text{and}\quad m_1-m_n<m+n,
\end{equation}
and let $\xi (\mathbf{m})$ be the associated solution of Eq.
\eqref{beq} detailed in Lemma \ref{sol:lem}. It follows from Lemmas
\ref{pauli:lem} and \ref{dist:lem} that
\begin{equation}\label{xeq}
 \xi_1(\mathbf{m})>\xi_2(\mathbf{m})>\cdots
> \xi_n(\mathbf{m})\quad \text{and}\quad
\xi_1(\mathbf{m})-\xi_n(\mathbf{m})<2\pi .
\end{equation}
\end{subequations}
Let us define
\begin{equation}\label{proj}
\boldsymbol{\mu}:= \mathbf{m}-\frac{1}{n}\langle
\mathbf{m},\mathbf{e}\rangle \,\mathbf{e}-\boldsymbol{\rho},\qquad
\boldsymbol{\xi}_{\boldsymbol{\mu}}:=\boldsymbol{\xi}(\mathbf{m})-\frac{1}{n}
\langle \boldsymbol{\xi}(\mathbf{m}),\mathbf{e}\rangle \,\mathbf{e},
\end{equation}
where (recall) $\mathbf{e}=\mathbf{e}_1+\cdots+\mathbf{e}_n$ and
$\boldsymbol{\rho}=\boldsymbol{\omega}_1+\cdots
+\boldsymbol{\omega}_{n-1}$. In other words, $\boldsymbol{\mu}$ is
the orthogonal projection of $\mathbf{m}$ onto the center-of-mass
hyperplane $\boldsymbol{E}$ \eqref{mass-plane} translated by
$-\boldsymbol{\rho}$ and $\boldsymbol{\xi}_{\boldsymbol{\mu}}$ is
the orthogonal projection of $\boldsymbol{\xi}(\mathbf{m})$ onto
$\boldsymbol{E}$. The inequalities in Eqs. \eqref{meq} and
\eqref{xeq} ensure that $\boldsymbol{\mu}\in\mathcal{P}^{(m)}$
\eqref{pm} and that $\boldsymbol{\xi}_{\boldsymbol{\mu}}\in 2\pi
\text{Int} (\boldsymbol{A})$ (cf. Eq. \eqref{alcove2}),
respectively. It is furthermore clear that by varying $\mathbf{m}$
we can reach any lattice point
$\boldsymbol{\mu}\in\mathcal{P}^{(m)}$. Indeed, for
$\boldsymbol{\mu}=k_1\boldsymbol{\omega}_1+\cdots
+k_{n-1}\boldsymbol{\omega}_{n-1}$ with $k_j\in\mathbb{Z}_{\geq 0}$
and $k_1+\cdots +k_{n-1}\leq m$ we may pick the components of
$\mathbf{m}$ equal to $m_j=\langle
\boldsymbol{\mu}+\boldsymbol{\rho},\boldsymbol{\alpha}_j+\cdots
+\boldsymbol{\alpha}_{n-1}\rangle =k_j+\cdots +k_{n-1}+n-j$,
$j=1,\ldots ,n$. It is not difficult to check that the assignment
$\boldsymbol{\mu}\to\boldsymbol{\xi}_{\boldsymbol{\mu}}$ is indeed
well-defined (i.e.
$\boldsymbol{\mu}=\boldsymbol{\mu}^\prime\Rightarrow
\boldsymbol{\xi}_{\boldsymbol{\mu}}=
\boldsymbol{\xi}_{\boldsymbol{\mu}^\prime}$) and one-to-one (i.e.
$\boldsymbol{\xi}_{\boldsymbol{\mu}}=
\boldsymbol{\xi}_{\boldsymbol{\mu}^\prime}\Rightarrow
\boldsymbol{\mu}=\boldsymbol{\mu}^\prime$). Indeed, one has that
\begin{eqnarray*}
\boldsymbol{\xi}_{\boldsymbol{\mu}}=
\boldsymbol{\xi}_{\boldsymbol{\mu}^\prime}
&\Longleftrightarrow &
\boldsymbol{\xi}(\mathbf{m})-\boldsymbol{\xi}(\mathbf{m}^\prime)\in
\mathbb{R}\mathbf{e} \\
&\stackrel{\text{Eq.}\, \eqref{beq}}{\Longleftrightarrow }&
\boldsymbol{\xi}(\mathbf{m})-\boldsymbol{\xi}(\mathbf{m}^\prime)\in
2\pi\mathbb{Z}\mathbf{e}\\
&\stackrel{\text{Eq.}\, \eqref{beq}}{\Longleftrightarrow } &
\mathbf{m}-\mathbf{m}^\prime\in
\mathbb{Z}\mathbf{e} \\
&\Longleftrightarrow & \boldsymbol{\mu}=\boldsymbol{\mu}^\prime .
\end{eqnarray*}
Since it is obvious that $\boldsymbol{\xi}_{\boldsymbol{\mu}}$
inherits from $\boldsymbol{\xi}(\mathbf{m})$ the smooth dependence
on the boundary parameter $t$ and the property that its components
solve the system in Eq. \eqref{beqp} (and thus the Bethe system in
Eq. \eqref{beqc}), this proves Theorem \ref{bethe-sol:thm} up to
Property {\em (ii)}. It remains to check Property {\em (iii)}, which
states that for $t=0$ the Bethe vectors are given by
$\boldsymbol{\xi}_{\boldsymbol{\mu}}=
\frac{2\pi}{n+m}(\boldsymbol{\rho}+\boldsymbol{\mu})$,
$\boldsymbol{\mu}\in\mathcal{P}^{(m)}$. To this end we simply
observe that Lemma \ref{pauli:lem} implies that for $t=0$
\begin{equation*}
 \xi_j(\mathbf{m})-\xi_{k}(\mathbf{m}) =
\frac{2\pi}{n+m}(m_j-m_{k}),
\end{equation*}
whence the statement follows by varying $\mathbf{m}$ subject to the
constraints in Eq. \eqref{meq} and projecting onto the
center-of-mass plane with the aid of Eq. \eqref{proj}.

\section{Diagonalization}\label{sec5}
In this section we will combine the results of Sections
\ref{sec2}--\ref{sec4} to arrive at an orthogonal basis for the
Hilbert space
$\mathcal{H}^{(m)}=\ell^2(\mathcal{P}^{(m)},\Delta^{(m)})$,
consisting of a complete set of joint eigenfunctions for the
(commuting) Laplace operators $L^{(m)}_1,\ldots ,L^{(m)}_{n-1}$.

\subsection{Spectrum and Eigenfunctions}
The following theorem provides the eigenfunctions of our Laplace
operators in terms of Hall-Littlewood polynomials specialized at the
Bethe vectors $\boldsymbol{\xi}_{\boldsymbol{\mu}}$,
$\boldsymbol{\mu}\in\mathcal{P}^{(m)}$.

\begin{theorem}[Spectrum and Eigenfunctions]\label{sev:thm}
For special values of the spectral parameter, given by the Bethe
vectors $\boldsymbol{\xi}_{\boldsymbol{\mu}}$,
$\boldsymbol{\mu}\in\mathcal{P}^{(m)}$ in Theorem
\ref{bethe-sol:thm}, the Bethe wave function
$\Psi_{\boldsymbol{\lambda}}(\boldsymbol{\xi})$ \eqref{hwf}
constitutes an eigenfunction of the Laplace operator $L_k^{(m)}$
\eqref{lapk}, \eqref{bconv}, i.e. for any $k\in\{ 1,\ldots ,n-1\}$
\begin{subequations}
\begin{equation}
L_k^{(m)}\,\Psi(\boldsymbol{\xi}_{\boldsymbol{\mu}})= E_k
(\boldsymbol{\xi}_{\boldsymbol{\mu}})\,
\Psi(\boldsymbol{\xi}_{\boldsymbol{\mu}}),
\end{equation}
where the eigenvalue is of the form
\begin{equation}\label{elv}
E_k(\boldsymbol{\xi})=
\sum_{\boldsymbol{\nu}\in\mathcal{S}_n(\boldsymbol{\omega}_k)}
\exp (i\langle \boldsymbol{\nu},\boldsymbol{\xi}\rangle )
\end{equation}
\end{subequations}
(and $\Psi(\boldsymbol{\xi}_{\boldsymbol{\mu}})\neq 0$).
\end{theorem}
\begin{proof}
Clearly the Hall-Littlewood polynomials
$\Psi_{\boldsymbol{\lambda}}(\boldsymbol{\xi})$  \eqref{hwf} satisfy
the identity
$\sum_{\boldsymbol{\nu}\in\mathcal{S}_n(\boldsymbol{\omega}_k)}
\Psi_{\boldsymbol{\lambda}+\boldsymbol{\nu}}(\boldsymbol{\xi})=
E_k(\boldsymbol{\xi})\Psi_{\boldsymbol{\lambda}}(\boldsymbol{\xi})$
(because all of the plane waves
$\psi_{\boldsymbol{\lambda}}(\boldsymbol{\xi}_\sigma)=\exp (i\langle
\boldsymbol{\lambda},\boldsymbol{\xi}_\sigma\rangle)$, $\sigma\in
\mathcal{S}_n$ do so). Moreover, since the specialized
Hall-Littlewood polynomials
$\Psi_{\boldsymbol{\lambda}}(\boldsymbol{\xi}_{\boldsymbol{\mu}})$,
$\boldsymbol{\mu}\in\mathcal{P}^{(m)}$ also satisfy the boundary
convention in Eq. \eqref{bconv} in view of Propositions
\ref{bref:prp}, \ref{bwf:prp}, \ref{bs:prp} and Theorem
\ref{bethe-sol:thm}, the stated eigenvalue equation follows. It
remains to check that
$\Psi_{\boldsymbol{\lambda}}(\boldsymbol{\xi}_{\boldsymbol{\mu}})$
does not vanish identically. For this purpose it is enough to
observe that for $\boldsymbol{\lambda}=\boldsymbol{0}$:
\begin{equation*}
\Psi_{\boldsymbol{0}} (\boldsymbol{\xi}) =  \sum_{\sigma\in
\mathcal{S}_n} \prod_{\boldsymbol{\alpha}\in \mathbf{R}^+}
\frac{1-t\, e^{-i\langle
\boldsymbol{\alpha},\boldsymbol{\xi}_\sigma\rangle} }{1-e^{-i\langle
\boldsymbol{\alpha},\boldsymbol{\xi}_\sigma\rangle} }
\stackrel{(i)}{=} \sum_{\sigma\in \mathcal{S}_n} t^{\ell (\sigma)}
\stackrel{(ii)}{=}\prod_{\boldsymbol{\alpha}\in\mathbf{R}^+}
\frac{1-t^{1+\langle \boldsymbol{\rho},\boldsymbol{\alpha}
\rangle}}{1-t^{\langle \boldsymbol{\rho},\boldsymbol{\alpha}
\rangle}} ,
\end{equation*}
where we have used {\em (i)} a rational function identity and {\em
(ii)} a product formula for the Poincar\'e series of the permutation
group that are both due to Macdonald \cite{mac:poincare} (cf.
Theorem (2.8) and Corollary (2.5), respectively). It is clear from
the product formula on the r.h.s. that $\Psi_{\boldsymbol{0}}
(\boldsymbol{\xi})>0$ for $-1<t<1$, whence
$\Psi_{\boldsymbol{\lambda}}(\boldsymbol{\xi}_{\boldsymbol{\mu}})$
indeed constitutes a true (i.e. nonzero) eigenfunction in
$\mathcal{H}^{(m)}$.
\end{proof}

\subsection{Orthogonality and Completeness}
Theorem \ref{sev:thm} provides as many eigenfunctions as the
dimension of the Hilbert space (indeed, $\dim(\mathcal{H}^{(m)})=\#
\mathcal{P}^{(m)}=\binom{n+m-1}{m}$). The following theorem confirms
our expectation that these eigenfunctions actually form an
orthogonal basis for the Hilbert space in question. Alternatively,
one may think of this theorem as describing a novel system of
discrete (dual) orthogonality relations for the Hall-Littlewood
polynomials.

\begin{theorem}[Orthogonality and Completeness]\label{oc:thm}
The Bethe wave functions
\begin{subequations}
\begin{equation}
\Psi (\boldsymbol{\xi}_{\boldsymbol{\mu}}),\qquad
\boldsymbol{\mu}\in\mathcal{P}^{(m)}
\end{equation}
 constitute an orthogonal basis
of $\mathcal{H}^{(m)}$:
\begin{equation}
\forall
\boldsymbol{\mu},\boldsymbol{\mu}^\prime\in\mathcal{P}^{(m)}: \qquad
 \langle \Psi (\boldsymbol{\xi}_{\boldsymbol{\mu}}), \Psi
(\boldsymbol{\xi}_{\boldsymbol{\mu}^\prime})\rangle^{(m)}=
\begin{cases}
0 &\text{if}\ \boldsymbol{\mu}\neq \boldsymbol{\mu}^\prime ,\\
>0 &\text{if}\ \boldsymbol{\mu}= \boldsymbol{\mu}^\prime .
\end{cases}
\end{equation}
\end{subequations}
\end{theorem}
\begin{proof}
Since $L_k^{(m)}$ and $L_{n-k}^{(m)}$ are each others adjoints in
$\mathcal{H}^{(m)}$ by Proposition \ref{adjoint:prp}, it is clear
that $ \langle L_k^{(m)} \Psi (\boldsymbol{\xi}_{\boldsymbol{\mu}}),
\Psi (\boldsymbol{\xi}_{\boldsymbol{\mu}^\prime})\rangle^{(m)} =
\langle \Psi (\boldsymbol{\xi}_{\boldsymbol{\mu}}),
L_{n-k}^{(m)}\Psi
(\boldsymbol{\xi}_{\boldsymbol{\mu}^\prime})\rangle^{(m)} $. By
applying Theorem \ref{sev:thm} and using the fact that
$E_k(\boldsymbol{\xi})=\overline{E_{n-k}(\boldsymbol{\xi})}$, this
equality is readily rewritten in the form
\begin{equation}\label{vanish}
\bigl( E_k(\boldsymbol{\xi}_{\boldsymbol{\mu}})-
E_k(\boldsymbol{\xi}_{\boldsymbol{\mu}^\prime}) \bigr) \langle \Psi
(\boldsymbol{\xi}_{\boldsymbol{\mu}}), \Psi
(\boldsymbol{\xi}_{\boldsymbol{\mu}^\prime})\rangle^{(m)}=0.
\end{equation}
Theorem \ref{bethe-sol:thm} now guarantees that for
$\boldsymbol{\mu}\neq \boldsymbol{\mu}^\prime$ the associated Bethe
vectors $\boldsymbol{\xi}_{\boldsymbol{\mu}}$ and
$\boldsymbol{\xi}_{\boldsymbol{\mu}^\prime}$ are distinct in $2\pi
\text{Int} (\boldsymbol{A})$. Moreover, since the elementary
symmetric polynomials $E_1(\boldsymbol{\xi}),\ldots,
E_{n-1}(\boldsymbol{\xi})$ separate the points of $2\pi \text{Int}
(\boldsymbol{A})$ (as they generate the full algebra of
trigonometric polynomials on $2\pi \boldsymbol{A}$ spanned by the
$\mathcal{S}_n$-invariant Fourier basis
$\sum_{\boldsymbol{\mu}\in\mathcal{S}_n(\boldsymbol{\lambda})} \exp
(i\langle \boldsymbol{\mu},\boldsymbol{\xi}\rangle)$,
$\boldsymbol{\lambda}\in \mathcal{P}$), this implies that in this
situation $E_k(\boldsymbol{\xi}_{\boldsymbol{\mu}})\neq
E_k(\boldsymbol{\xi}_{\boldsymbol{\mu}^\prime})$ for a certain value
of $k\in \{1,\ldots ,n-1\}$. We thus conclude from Eq.
\eqref{vanish} that the inner product $\langle \Psi
(\boldsymbol{\xi}_{\boldsymbol{\mu}}), \Psi
(\boldsymbol{\xi}_{\boldsymbol{\mu}^\prime})\rangle^{(m)}$ must
vanish if $\boldsymbol{\mu}\neq \boldsymbol{\mu}^\prime$. Finally,
for $\boldsymbol{\mu}=\boldsymbol{\mu}^\prime$ the inner product
yields the squared norm of the Bethe wave function $\Psi
(\boldsymbol{\xi}_{\boldsymbol{\mu}})$ in $\mathcal{H}^{(m)}$, which
is positive as $\Psi (\boldsymbol{\xi}_{\boldsymbol{\mu}})\neq 0$ by
(the proof of) Theorem \ref{sev:thm}.
\end{proof}

\subsection{Integrability}
From the previous results it is seen that our Laplace operators
model a finite-dimensional quantum system that is integrable in the
following sense.

\begin{theorem}[Integrability]\label{int:thm}
The Laplacians $L_1^{(m)},\ldots ,L_{n-1}^{(m)}$ \eqref{lapk},
\eqref{bconv} constitute $n-1$ ($=\dim (\boldsymbol{E})$) mutually
commuting operators in the Hilbert space $\mathcal{H}^{(m)}$.
Furthermore, any operator $L:\mathcal{H}^{(m)}\to\mathcal{H}^{(m)}$
that commutes with all of the Laplacians $L_1^{(m)},\ldots
,L_{n-1}^{(m)}$ lies in the polynomial algebra
$\mathbb{C}[L_1^{(m)},\ldots ,L_{n-1}^{(m)}]$.
\end{theorem}
\begin{proof}
The commutativity of $L_1^{(m)},\ldots ,L_{n-1}^{(m)}$ is immediate
from the fact that the operators are simultaneously diagonalized by
the basis $\Psi (\boldsymbol{\xi}_{\boldsymbol{\mu}})$,
$\boldsymbol{\mu}\in\mathcal{P}^{(m)}$ of $\mathcal{H}^{(m)}$ (cf.
Theorems \ref{sev:thm} and \ref{oc:thm}). The property that any
operator $L:\mathcal{H}^{(m)}\to\mathcal{H}^{(m)}$ that commutes
with $L_1^{(m)},\ldots ,L_{n-1}^{(m)}$ is necessarily algebraically
dependent of $L_1^{(m)},\ldots ,L_{n-1}^{(m)}$ hinges on the fact
that the eigenvalues
$E_1(\boldsymbol{\xi}_{\boldsymbol{\mu}}),\ldots
,E_{n-1}(\boldsymbol{\xi}_{\boldsymbol{\mu}})$ separate the elements
of the eigenbasis $\Psi (\boldsymbol{\xi}_{\boldsymbol{\mu}})$,
$\boldsymbol{\mu}\in\mathcal{P}^{(m)}$ (cf. also the proof of
Theorem \ref{oc:thm}). Indeed, it is immediate from this that $L$ is
diagonalized by $\Psi (\boldsymbol{\xi}_{\boldsymbol{\mu}})$,
$\boldsymbol{\mu}\in\mathcal{P}^{(m)}$. In other words, that there
exist a function $E_L:\{ \boldsymbol{\xi}_{\boldsymbol{\mu}}
\}_{\boldsymbol{\mu}\in\mathcal{P}^{(m)}}\to\mathbb{C}$ such that
\begin{subequations}
\begin{equation}\label{a}
L\Psi
(\boldsymbol{\xi}_{\boldsymbol{\mu}})=E_L(\boldsymbol{\xi}_{\boldsymbol{\mu}})
\Psi (\boldsymbol{\xi}_{\boldsymbol{\mu}}),\quad \forall
\boldsymbol{\mu}\in\mathcal{P}^{(m)}.
\end{equation}
Since the Bethe functions $\Psi
(\boldsymbol{\xi}_{\boldsymbol{\mu}})$,
$\boldsymbol{\mu}\in\mathcal{P}^{(m)}$ form an orthogonal basis of
$\mathcal{H}^{(m)}$, we have (by transposition) that the
Hall-Littlewood polynomials
$\Psi_{\boldsymbol{\lambda}}(\boldsymbol{\xi})$,
$\boldsymbol{\lambda}\in\mathcal{P}^{(m)}$ form a basis for the
space of complex functions on the spectral set $\{
\boldsymbol{\xi}_{\boldsymbol{\mu}}
\}_{\boldsymbol{\mu}\in\mathcal{P}^{(m)}}$ upon specialization. In
particular, there exist (unique) complex coefficients
$c_{\boldsymbol{\lambda}}$,
$\boldsymbol{\lambda}\in\mathcal{P}^{(m)}$ such that
\begin{equation}\label{b}
E_L(\boldsymbol{\xi}_{\boldsymbol{\mu}})=
\sum_{\boldsymbol{\lambda}\in\mathcal{P}^{(m)}}
c_{\boldsymbol{\lambda}}
\Psi_{\boldsymbol{\lambda}}(\boldsymbol{\xi}_{\boldsymbol{\mu}}),\quad
\forall\boldsymbol{\mu}\in \mathcal{P}^{(m)}.
\end{equation}
Furthermore, from the well-known property that the elementary
symmetric polynomials $E_1(\boldsymbol{\xi}),\ldots,
E_{n-1}(\boldsymbol{\xi})$ \eqref{elv} generate the space of
symmetric polynomials it is clear that there exist a polynomial
$P_L\in\mathbb{C}[E_1,\ldots ,E_{n-1}]$ such that
\begin{equation}\label{c}
 \sum_{\boldsymbol{\lambda}\in\mathcal{P}^{(m)}}
c_{\boldsymbol{\lambda}}
\Psi_{\boldsymbol{\lambda}}(\boldsymbol{\xi})=
P_L(E_1(\boldsymbol{\xi}),\ldots ,E_{n-1}(\boldsymbol{\xi})) .
\end{equation}
\end{subequations}
It follows from Eqs. \eqref{a}--\eqref{c} and Theorem \ref{sev:thm}
that the operators $L$ and $P_L(L_1^{(m)},\ldots ,L_{n-1}^{(m)})$
coincide on the basis $\Psi (\boldsymbol{\xi}_{\boldsymbol{\mu}})$,
$\boldsymbol{\mu}\in\mathcal{P}^{(m)}$. Hence, we conclude that
$L=P_L(L_1^{(m)},\ldots ,L_{n-1}^{(m)})\in
\mathbb{C}[L_1^{(m)},\ldots ,L_{n-1}^{(m)}]$.
\end{proof}

The Laplace operators $L_1^{(m)},\ldots ,L_{n-1}^{(m)}$ are not
self-adjoint in general in view of Proposition \ref{adjoint:prp}. As
a consequence, the spectrum in Theorem \ref{sev:thm} is generally
complex-valued. Within the commuting algebra
$\mathbb{C}[L_1^{(m)},\ldots ,L_{n-1}^{(m)}]$ there exist however
many operators that {\em are} self-adjoint. For example, the
alternative generators
\begin{subequations}
\begin{eqnarray}
L_{R,k}^{(m)}&:=& \frac{1}{2}\bigl(
L_k^{(m)}+L_{n-k}^{(m)}\bigr),\qquad k\in \{
1,\ldots ,[n/2] \} , \\
L_{I,k}^{(m)}&:=& \frac{1}{2i}\bigl(
L_k^{(m)}-L_{n-k}^{(m)}\bigr),\qquad k\in \{ 1,\ldots ,[(n-1)/2] \}
,
\end{eqnarray}
\end{subequations}
are self-adjoint and have real spectrum of the form
\begin{subequations}
\begin{eqnarray}
E_{R,k}(\boldsymbol{\xi}_{\boldsymbol{\mu}})&=&
\sum_{\boldsymbol{\nu}\in\mathcal{S}_n(\boldsymbol{\omega}_k)}\cos
( \langle
\boldsymbol{\nu},\boldsymbol{\xi}_{\boldsymbol{\mu}}\rangle ),
\qquad \boldsymbol{\mu}\in\mathcal{P}^{(m)}, \\
E_{I,k}(\boldsymbol{\xi}_{\boldsymbol{\mu}})&=&
\sum_{\boldsymbol{\nu}\in\mathcal{S}_n(\boldsymbol{\omega}_k)}\sin
( \langle
\boldsymbol{\nu},\boldsymbol{\xi}_{\boldsymbol{\mu}}\rangle ),
\qquad \boldsymbol{\mu}\in\mathcal{P}^{(m)},
\end{eqnarray}
\end{subequations}
respectively. The real subalgebra $\mathbb{R}[L_{R,1}^{(m)},\ldots
,L_{R,[n/2]}^{(m)},L_{I,1}^{(m)},\ldots ,L_{I,[(n-1)2]}^{(m)}]$
consists of all operators $L:\mathcal{H}^{(m)}\to\mathcal{H}^{(m)}$
such that {\em (i)} $L$ commutes with all of the Laplacians
$L_1^{(m)},\ldots ,L_{n-1}^{(m)}$ and {\em (ii)} $L$ is
self-adjoint.

One of the simplest positive operators in this real subalgebra is
given by
\begin{equation}
H^{(m)}:=n\text{Id}-L_{R,1}^{(m)}.
\end{equation}
 In standard coordinates the explicit action of this operator on
an arbitrary wave function $\psi\in\mathcal{H}^{(m)}$ is of the form
(cf. Proposition \ref{action:prp})
\begin{subequations}
\begin{equation}\label{ham1}
(H^{(m)}\psi)_{\boldsymbol{\lambda}} =n\psi_{\boldsymbol{\lambda}}
-\frac{1}{2} \sum_{\begin{subarray}{c} 1\leq j\leq n \\
\boldsymbol{\lambda}+\boldsymbol{\nu}_j\in\mathcal{P}^{(m)}\end{subarray}}
V_{j,\boldsymbol{\lambda}}^+\,\psi_{\boldsymbol{\lambda}+\boldsymbol{\nu}_j}
-\frac{1}{2} \sum_{\begin{subarray}{c} 1\leq j\leq n \\
\boldsymbol{\lambda}-\boldsymbol{\nu}_j\in\mathcal{P}^{(m)}\end{subarray}}
V_{j,\boldsymbol{\lambda}}^-\,\psi_{\boldsymbol{\lambda}-\boldsymbol{\nu}_j}
 ,
\end{equation}
where
\begin{eqnarray}
V_{j,\boldsymbol{\lambda}}^+ &=& \prod_{\begin{subarray}{c} j<k\leq n\\
\lambda_k=\lambda_j
\end{subarray}}  \frac{1-t^{1+k-j}}{1-t^{k-j}}
\prod_{\begin{subarray}{c} 1\leq k <j\\
\lambda_k=\lambda_j+m
\end{subarray}}  \frac{1-t^{1+n+k-j}}{1-t^{n+k-j}}  ,\\
V_{j,\boldsymbol{\lambda}}^- &=&
\prod_{\begin{subarray}{c}1\leq k <j\\
\lambda_k=\lambda_j
\end{subarray}}  \frac{1-t^{1+j-k}}{1-t^{j-k}}
\prod_{\begin{subarray}{c} j<k \leq n\\
\lambda_k=\lambda_j-m
\end{subarray}}  \frac{1-t^{1+n+j-k}}{1-t^{n+j-k}} , \label{ham2}
\end{eqnarray}
and $\boldsymbol{\nu}_j=\mathbf{e}_j-(\mathbf{e}_1+\cdots
+\mathbf{e}_n)/n$, $j=1,\ldots ,n$ (so $\boldsymbol{\nu}_1,\ldots
,\boldsymbol{\nu}_n$ consist of the orthogonal projection of the
standard basis $\mathbf{e}_1,\ldots ,\mathbf{e}_n$ onto the
center-of-mass plane $\boldsymbol{E}$ \eqref{mass-plane}). The
spectrum of $H^{(m)}$ is built of positive eigenvalues
$E(\boldsymbol{\xi}_{\boldsymbol{\mu}})$,
$\boldsymbol{\mu}\in\mathcal{P}^{(m)}$ with
\begin{equation}\label{evham}
E(\boldsymbol{\xi})=\sum_{j=1}^n \bigl( 1-\cos (\xi_j )\bigr) .
\end{equation}
\end{subequations}
The operator $H^{(m)}$ \eqref{ham1}--\eqref{ham2} serves as the
Hamiltonian of our lattice $n$-particle model. Below we will verify
that in a continuum limit this lattice Hamiltonian tends formally to
the Hamiltonian of the $n$-particle delta Bose gas on the circle.

\section{Continuum Limit}\label{sec6}
In this final section we first review the solution of the eigenvalue
problem for the Laplacian in Eq. \eqref{ep}, with wave functions
supported inside the alcove $\boldsymbol{A}$ \eqref{alcove} subject
to repulsive boundary conditions of the form in Eqs. \eqref{wall1},
\eqref{wall2} (i.e. with $g>0$). Our formulation amounts to the
center-of-mass reduction of the seminal results due to Lieb and
Liniger \cite{lie-lin:exact} (Bethe wave functions), C.N. Yang and
C.P. Yang \cite{yan-yan:thermodynamics} (Bethe vectors), and Dorlas
\cite{dor:orthogonality} (orthogonality and completeness). Next we
will show how this solution of the eigenvalue problem for the
Laplacian in the alcove can be recovered from the corresponding
solution of our discrete lattice model via a continuum limit.

\subsection{Eigenfunctions}
In the notation of Section \ref{sec2} the eigenvalue problem in Eqs.
\eqref{ep}--\eqref{wall2} reads
\begin{subequations}
\begin{equation}\label{eva}
-\Delta \psi = E \psi,\qquad \mathbf{x}\in\boldsymbol{A} ,
\end{equation}
with
\begin{eqnarray}\label{ba}
&& \bigl( \langle \nabla_{\mathbf{x}}\psi
,\boldsymbol{\alpha}_0\rangle +g\psi \bigr)
|_{\mathbf{x}\in\boldsymbol{E}_0}=0 ,\\
&& \bigl( \langle \nabla_{\mathbf{x}}\psi
,\boldsymbol{\alpha}_j\rangle -g\psi \bigr)
|_{\mathbf{x}\in\boldsymbol{E}_j}=0,\quad j=1,\ldots ,n-1\label{bb}
\end{eqnarray}
\end{subequations}
(where $\nabla_{\mathbf{x}}$ refers to the gradient). Let us define
\begin{subequations}
\begin{eqnarray}
\boldsymbol{C} &:=&\{ \boldsymbol{\xi}\in\boldsymbol{E}\mid \langle
\boldsymbol{\xi},\boldsymbol{\alpha}_j\rangle >0,\; j=1,\ldots
,n-1\} ,\label{chamber} \\
\mathcal{P}^{(\infty)} &:=& \{ k_1\boldsymbol{\omega}_1+\cdots
+k_{n-1}\boldsymbol{\omega}_{n-1}\mid k_1,\ldots
,k_{n-1}\in\mathbb{Z}_{\geq 0} \} .\label{domw}
\end{eqnarray}
\end{subequations}

\begin{theorem}[Bethe Wave Functions
\cite{lie-lin:exact}]\label{lieb-lin:thm} The Bethe wave
function\footnote{This explicit form of the expressions for the
coefficients of the Bethe wave function is due to Gaudin
\cite{gau:boundary,gau:fonction}.}
\begin{subequations}
\begin{equation}\label{lieb-lin-wf}
\Psi^{(\infty )} (\mathbf{x},\boldsymbol{\xi})=
\sum_{\sigma\in\mathcal{S}_n} \Bigl( \prod_{\boldsymbol{\alpha}\in
\mathbf{R}^+} \frac{\langle \boldsymbol{\alpha}
,\boldsymbol{\xi}_\sigma \rangle -ig}{\langle \boldsymbol{\alpha}
,\boldsymbol{\xi}_\sigma\rangle} \Bigr) e^{i\langle
\mathbf{x},\boldsymbol{\xi}_\sigma\rangle} ,
\end{equation}
with the spectral parameter $\boldsymbol{\xi}\in \boldsymbol{C}$
\eqref{chamber} solving the Bethe system
\begin{equation}\label{bs0}
e^{i\langle \boldsymbol{\beta},\boldsymbol{\xi}\rangle}= \left(
\frac{ig+\langle\boldsymbol{\beta},\boldsymbol{\xi}\rangle}
{ig-\langle\boldsymbol{\beta},\boldsymbol{\xi}\rangle}\right)^2
\prod_{\begin{subarray}{c}  \boldsymbol{\alpha}\in \mathbf{R} \\
\langle \boldsymbol{\alpha},\boldsymbol{\beta}\rangle =1
\end{subarray}}
\frac{ig+\langle\boldsymbol{\alpha},\boldsymbol{\xi}\rangle}
{ig-\langle\boldsymbol{\alpha},\boldsymbol{\xi}\rangle},
\qquad \forall \boldsymbol{\beta}\in\mathbf{R},
\end{equation}
\end{subequations}
constitutes a solution to the eigenvalue problem in Eqs.
\eqref{eva}--\eqref{bb} corresponding to the eigenvalue
$E=E^{(\infty)} (\boldsymbol{\xi}):= \langle
\boldsymbol{\xi},\boldsymbol{\xi}\rangle $.
\end{theorem}

It is instructive to recall briefly the essence of the proof of Lieb
and Liniger in the present notation. Firstly, it is clear that the
linear combination of plane waves $\Psi^{(\infty
)}(\mathbf{x},\boldsymbol{\xi})$ constitutes an eigenfunction of
$-\Delta$ with eigenvalue $E^{(\infty)}(\boldsymbol{\xi})$. It
remains to check that the boundary conditions are also satisfied.
The boundary condition in Eq. \eqref{bb} is inferred by the
following computation for $\mathbf{x}\in \boldsymbol{E}_j$:
\begin{eqnarray*}
\lefteqn{\langle \nabla_{\mathbf{x}} \Psi^{(\infty )},
\boldsymbol{\alpha}_j\rangle } &&\\
&=& \sum_{\sigma \in \mathcal{S}_n}
\Bigl(\prod_{\boldsymbol{\alpha}\in\mathbf{R}^+}
\frac{\langle\boldsymbol{\alpha}
,\boldsymbol{\xi}_\sigma\rangle-ig }{\langle\boldsymbol{\alpha}
,\boldsymbol{\xi}_\sigma\rangle} \Bigr)
i\langle\boldsymbol{\alpha}_j ,\boldsymbol{\xi}_\sigma\rangle
e^{i\langle\mathbf{x}
,\boldsymbol{\xi}_\sigma\rangle}  \\
&=& \sum_{\sigma\in \mathcal{S}_n} (g+i\langle\boldsymbol{\alpha}_j
,\boldsymbol{\xi}_\sigma\rangle )
\Bigl(\prod_{\begin{subarray}{c}\boldsymbol{\alpha}\in\mathbf{R}^+\\
\boldsymbol{\alpha}\neq \boldsymbol{\alpha}_j\end{subarray}
}\frac{\langle\boldsymbol{\alpha} ,\boldsymbol{\xi}_\sigma\rangle-ig
}{\langle\boldsymbol{\alpha} ,\boldsymbol{\xi}_\sigma\rangle} \Bigr)
e^{i\langle\mathbf{x}
,\boldsymbol{\xi}_\sigma\rangle} \\
&\stackrel{(i)}{=}& g\sum_{\sigma\in \mathcal{S}_n}
\Bigl(\prod_{\begin{subarray}{c}\boldsymbol{\alpha}\in\mathbf{R}^+\\
\boldsymbol{\alpha}\neq \boldsymbol{\alpha}_j\end{subarray}
}\frac{\langle\boldsymbol{\alpha}
,\boldsymbol{\xi}_\sigma\rangle-ig}{\langle\boldsymbol{\alpha}
,\boldsymbol{\xi}_\sigma\rangle} \Bigr) e^{i\langle\mathbf{x}
,\boldsymbol{\xi}_\sigma\rangle} \\
&\stackrel{(ii)}{=}& g\sum_{\sigma\in \mathcal{S}_n} \Bigl(
1-\frac{ig}{\langle\boldsymbol{\alpha}_j
,\boldsymbol{\xi}_\sigma\rangle} \Bigr)
\Bigl(\prod_{\begin{subarray}{c}\boldsymbol{\alpha}\in\mathbf{R}^+\\
\boldsymbol{\alpha}\neq \boldsymbol{\alpha}_j\end{subarray}
}\frac{\langle\boldsymbol{\alpha} ,\boldsymbol{\xi}_\sigma\rangle-ig
}{\langle\boldsymbol{\alpha} ,\boldsymbol{\xi}_\sigma\rangle} \Bigr)
e^{i\langle\mathbf{x}
,\boldsymbol{\xi}_\sigma\rangle} \\
&=& g\sum_{\sigma\in \mathcal{S}_n}
\Bigl(\prod_{\boldsymbol{\alpha}\in\mathbf{R}^+
}\frac{\langle\boldsymbol{\alpha} ,\boldsymbol{\xi}_\sigma\rangle-ig
}{\langle\boldsymbol{\alpha} ,\boldsymbol{\xi}_\sigma\rangle} \Bigr)
e^{i\langle\mathbf{x}
,\boldsymbol{\xi}_\sigma\rangle} \\
&=& g\Psi^{(\infty )} ,
\end{eqnarray*}
where in Steps $(i)$ and $(ii)$ one exploits that
$\prod_{\begin{subarray}{c}\boldsymbol{\alpha}\in\mathbf{R}^+\\
\boldsymbol{\alpha}\neq \boldsymbol{\alpha}_j\end{subarray}
}\frac{\langle\boldsymbol{\alpha} ,\boldsymbol{\xi}_\sigma\rangle-ig
}{\langle\boldsymbol{\alpha} ,\boldsymbol{\xi}_\sigma\rangle}$ and
$\langle\boldsymbol{\alpha}_j ,\boldsymbol{\xi}_\sigma\rangle$ are
symmetric and skew-symmetric, respectively, with respect to the
action of $r_j$ on $\boldsymbol{\xi}_\sigma$, combined with the
symmetry $\langle\mathbf{x}
,r_j(\boldsymbol{\xi}_\sigma)\rangle=\langle\mathbf{x}
,\boldsymbol{\xi}_\sigma\rangle$ (since $r_j(\mathbf{x})=\mathbf{x}$
if $\mathbf{x}\in\boldsymbol{E}_j$). Finally, the boundary condition
in Eq. \eqref{ba} requires that for $\mathbf{x}\in \boldsymbol{E}_0$
\begin{equation*}
 \sum_{\sigma\in\mathcal{S}_n}
 (g+i\langle \boldsymbol{\xi}_\sigma ,\boldsymbol{\alpha}_0\rangle )
\Bigl( \prod_{\boldsymbol{\alpha}\in \mathbf{R}^+} \frac{\langle
\boldsymbol{\alpha} ,\boldsymbol{\xi}_\sigma \rangle -ig}{\langle
\boldsymbol{\alpha} ,\boldsymbol{\xi}_\sigma\rangle} \Bigr)
e^{i\langle \mathbf{x},\boldsymbol{\xi}_\sigma\rangle} =0.
\end{equation*}
Manipulations similar to those in the proof of Proposition
\ref{bs:prp} reveal that this relation holds when the spectral
parameter solves the Bethe system in Eq. \eqref{bs0}.

\begin{theorem}[Bethe Vectors
\cite{yan-yan:thermodynamics}]\label{yang-yang:thm} Let $g>0$. For
each $\boldsymbol{\mu}\in \mathcal{P}^{(\infty)}$ \eqref{domw} there
exists a (unique) Bethe vector
$\boldsymbol{\xi}_{\boldsymbol{\mu}}\in \boldsymbol{C}$
\eqref{chamber} such that $\boldsymbol{\xi}_{\boldsymbol{\mu}}$
satisfies the system in Eq. \eqref{bs0}. Moreover, these Bethe
vectors have the following properties:
\begin{itemize}
\item[{\em (i)}]
$\boldsymbol{\xi}_{\boldsymbol{\mu}^\prime}=\boldsymbol{\xi}_{\boldsymbol{\mu}}$
if and only if $\boldsymbol{\mu}^\prime=\boldsymbol{\mu}$,
\item[{\em (ii)}] $\boldsymbol{\xi}_{\boldsymbol{\mu}}$ depends
smoothly on the boundary parameter $g>0$,
\item[{\em (iii)}]
$\boldsymbol{\xi}_{\boldsymbol{\mu}}\to
2\pi(\boldsymbol{\rho}+\boldsymbol{\mu})$ for $g\to +\infty$.
\end{itemize}
\end{theorem}

In standard coordinates the Bethe system in Eq. \eqref{bs0} reads
\begin{equation}
e^{i(\xi_j-\xi_k)}= \prod_{\begin{subarray}{c} 1\leq \ell\leq n
\\ \ell \neq j\end{subarray}}
\frac{ig+\xi_j-\xi_\ell}{ig-\xi_j+\xi_\ell}
\prod_{\begin{subarray}{c} 1\leq \ell\leq n
\\ \ell \neq k\end{subarray}}
\frac{ig+\xi_\ell-\xi_k}{ig-\xi_\ell+\xi_k},
\end{equation}
for $1\leq j\neq k\leq n$, or equivalently (upon exploiting the
translational invariance)
\begin{equation}
e^{i\xi_j}= \prod_{\begin{subarray}{c} 1\leq \ell\leq n
\\ \ell \neq j\end{subarray}}
\frac{ig+\xi_j-\xi_\ell}{ig-\xi_j+\xi_\ell}, \qquad j=1,\ldots ,n.
\end{equation}
In the additive form the latter system becomes
\begin{equation}\label{beq0}
\xi_j+\sum_{\begin{subarray}{c} 1\leq \ell\leq n
\\ \ell \neq j\end{subarray}} \theta^{(\infty )} (\xi_j-\xi_\ell) = 2\pi m_j,
\qquad j=1,\ldots ,n,
\end{equation}
with $\mathbf{m}=(m_1,\ldots ,m_n)\in \mathbb{Z}^n$ and
\begin{subequations}
\begin{eqnarray}
\theta^{(\infty )} (x) &= & 2g\int_0^x (x^2+g^2)^{-1} \text{d}x
\\
&=& 2\arctan  \bigl(\frac{x}{g}\bigr) \\
&=& i\log \left( \frac{ig+x}{ig-x} \right) .\label{theta03}
\end{eqnarray}
\end{subequations}
It was shown in \cite{yan-yan:thermodynamics} that for any
$\mathbf{m}\in\mathbb{Z}^n$ the Bethe system in Eqs.
\eqref{beq0}--\eqref{theta03} has a unique solution
$\boldsymbol{\xi} (\mathbf{m})$ given by the unique global minimum
of the strictly convex function
\begin{equation}\label{Vinf}
V^{(\infty )}(\xi_1,\ldots ,\xi_n):= \frac{1}{2}\sum_{j=1}^n \xi_j^2
+ \frac{1}{2} \sum_{j,k =1}^n \Theta^{(\infty )} (\xi_j-\xi_k )
-2\pi \sum_{j=1}^n m_j\xi_j ,
\end{equation}
with $\Theta^{(\infty )}(x):=\int_0^x \theta^{(\infty
)}(x)\text{d}x$. By projecting the solutions
$\boldsymbol{\xi}(\mathbf{m})$, corresponding to vectors
$\mathbf{m}\in\mathbb{Z}^n$ with $m_1>m_2>\cdots >m_n$, orthogonally
onto the center-of-mass plane $\boldsymbol{E}$ the statements of
Theorem \ref{yang-yang:thm} readily follow (cf. also Section
\ref{sec4}).

\begin{theorem}[Orthogonality and Completeness
\cite{dor:orthogonality}]\label{dorlas:thm} The Bethe wave functions
$\Psi^{(\infty )}(\mathbf{x},\boldsymbol{\xi}_{\boldsymbol{\mu}})$,
$\boldsymbol{\mu}\in\mathcal{P}^{(\infty)}$ form an orthogonal basis
for the Hilbert space $\mathcal{H}^{(\infty
)}:=L^2(\boldsymbol{A},\text{d}\mathbf{x})$ (with inner product
$\langle \phi,\psi\rangle^{(\infty)}:= \int_{\boldsymbol{A}} \phi
(\mathbf{x})\overline {\psi (\mathbf{x})}\text{d}\mathbf{x}$), i.e.
\begin{subequations}
\begin{equation}
\forall
\boldsymbol{\mu},\boldsymbol{\mu}^\prime\in\mathcal{P}^{(\infty)}:
\qquad
 \langle \Psi^{(\infty )} (\boldsymbol{\xi}_{\boldsymbol{\mu}}), \Psi^{(\infty )}
(\boldsymbol{\xi}_{\boldsymbol{\mu}^\prime})\rangle^{(\infty)} =
\begin{cases}
0 &\text{if}\ \boldsymbol{\mu}\neq \boldsymbol{\mu}^\prime \\
>0 &\text{if}\ \boldsymbol{\mu}= \boldsymbol{\mu}^\prime
\end{cases}
\end{equation}
and
\begin{equation}
\langle \phi, \Psi^{(\infty )}
(\boldsymbol{\xi}_{\boldsymbol{\mu}})\rangle =0,\
\forall\boldsymbol{\mu}\in\mathcal{P}^{(\infty)} \Longrightarrow
\phi =0.
\end{equation}
\end{subequations}
\end{theorem}

Below we will infer that this center-of-mass reduction of Dorlas'
orthogonality relations can be recovered via a continuum limit from
the corresponding results pertaining to the discrete lattice model
in Section \ref{sec5}. It was moreover shown by Dorlas that the
orthogonality of the Bethe wave functions for the repulsive delta
Boson gas implies their completeness \cite[Section
3]{dor:orthogonality}. In other words, the completeness in Theorem
\ref{dorlas:thm} follows from the orthogonality (upon a cosmetic
adaptation of Dorlas' arguments to our center-of-mass situation).

\subsection{Orthogonality}
In order to perform the continuum limit let us from now on rescale
the coupling parameter $t$ putting
\begin{equation}
t=e^{-g/m},\qquad g>0.
\end{equation}
For any $\mathbf{x}$ in (the closure of) $\boldsymbol{C}$
\eqref{chamber} we define an integral approximation
$[\mathbf{x}]\in\mathcal{P}^{(\infty)}$ \eqref{domw} of the form
\begin{equation}
[\mathbf{x}]:= [\langle \mathbf{x},\boldsymbol{\alpha}_1\rangle ]
\boldsymbol{\omega}_1+\cdots +[\langle
\mathbf{x},\boldsymbol{\alpha}_{n-1}\rangle ]
\boldsymbol{\omega}_{n-1} ,
\end{equation}
where $[x]$ denotes the integral part of $x\in \mathbb{R}_{\geq 0}$
obtained through truncation. With these notations we are in the
position to embed the Hilbert space of lattice functions
$\mathcal{H}^{(m)}=\ell^2(\mathcal{P}^{(m)},\Delta^{(m)})$ into
$L^2(\boldsymbol{C},\text{d}x)$ by means of a linear injection
$J^{(m)}:\mathcal{H}^{(m)}\to L^2(\boldsymbol{C},\text{d}x)$ that
associates to a lattice function
$\phi:\mathcal{P}^{(m)}\to\mathbb{C}$ a staircase function
$J^{(m)}(\phi):\boldsymbol{C}\to\mathbb{C}$ of the form
\begin{equation}
(J^{(m)}\phi)(\mathbf{x}):=
\begin{cases}
\sqrt{\Delta_{[m\mathbf{x}]}^{(m)}} \phi_{[m\mathbf{x}]}
&\text{for}\ [m\mathbf{x}]\in\mathcal{P}^{(m)} ,\\
0&\text{for}\ [m\mathbf{x}]\not\in\mathcal{P}^{(m)} .
\end{cases}
\end{equation}
It is not difficult to see that the staircase function
$J^{(m)}(\phi)$ has support on a bounded domain inside the dilated
alcove $(1+\frac{n}{m})\boldsymbol{A}$.  This support shrinks
towards (a subset of) $\boldsymbol{A}$ for $m\to\infty$. It is also
not difficult to deduce from this definition that $\forall
\phi,\psi\in\mathcal{H}^{(m)}$
\begin{equation}\label{ipr}
\int_{\boldsymbol{C}} (J^{(m)}\phi)(\mathbf{x})\overline{
(J^{(m)}\psi) (\mathbf{x})}\text{d}\mathbf{x} =
c_{n,m}\sum_{\boldsymbol{\lambda}\in\mathcal{P}^{(m)}}
\phi_{\boldsymbol{\lambda}} \overline{ \psi_{\boldsymbol{\lambda}}}
\Delta^{(m )}_{\boldsymbol{\lambda}} ,
\end{equation}
where
$c_{n,m}=\text{Vol}(\boldsymbol{\omega}_1,\ldots,\boldsymbol{\omega}_{n-1})/m^{n-1}
= 1/(m^{n-1}\sqrt{n})$. Let $\Psi^{(m)}
(\mathbf{x},\boldsymbol{\xi})$ be the staircase embedding of the
Hall-Littlewood polynomial
$\Psi_{\boldsymbol{\lambda}}(\boldsymbol{\xi})$  \eqref{hwf}
\begin{eqnarray}
\Psi^{(m)} (\mathbf{x},\boldsymbol{\xi}) &:=& (J^{(m)}
\Psi(\boldsymbol{\xi}))(\mathbf{x}) \label{staircase} \\
&= &\sqrt{\Delta_{[m\mathbf{x}]}^{(m)}}
\Psi_{[m\mathbf{x}]}(\boldsymbol{\xi}), \nonumber \\
&=& \sqrt{\Delta_{[m\mathbf{x}]}^{(m)}}\sum_{\sigma\in
\mathcal{S}_n}\Bigl( \prod_{\boldsymbol{\alpha}\in \mathbf{R}^+}
\frac{1-e^{-g/m}\, e^{-i\langle
\boldsymbol{\alpha},\boldsymbol{\xi}_\sigma\rangle} }{1-e^{-i\langle
\boldsymbol{\alpha},\boldsymbol{\xi}_\sigma\rangle} }\Bigr)
e^{i\langle [m\mathbf{x}] ,\boldsymbol{\xi}_\sigma \rangle} .
\nonumber
\end{eqnarray}

The following lemma states that, for $m\to \infty$, the rescaled
staircase function $\Psi^{(m)}
(\mathbf{x},\frac{1}{m}\boldsymbol{\xi})$ \eqref{staircase}
converges pointwise to the Lieb-Liniger Bethe wave function
$\Psi^{(\infty )} (\mathbf{x},\boldsymbol{\xi})$ \eqref{lieb-lin-wf}
when $\mathbf{x}$ lies in the interior of $\boldsymbol{A}$ and to
zero when $\mathbf{x}$ lies outside $\boldsymbol{A}$.

\begin{lemma}\label{wflim:lem}
For any $\boldsymbol{\xi}\in\boldsymbol{C}$, one has that
\begin{equation}
\lim_{m\to\infty} \Psi^{(m)} (\mathbf{x},{\textstyle
\frac{1}{m}}\boldsymbol{\xi}) =
\begin{cases}
\Psi^{(\infty )} (\mathbf{x},\boldsymbol{\xi}) &\text{if}\
\mathbf{x}\in \text{Int}(\boldsymbol{A}) ,\\
0&\text{if}\ \mathbf{x}\in\boldsymbol{C}\setminus\boldsymbol{A}.
\end{cases}
\end{equation}
\end{lemma}
\begin{proof}
The lemma readily follows from the explicit expression of the
staircase wave function on the third line of Eq. \eqref{staircase},
together with the observation that $\lim_{m\to\infty}
\Delta_{[m\mathbf{x}]}^{(m)}= 1$ if
$\mathbf{x}\in\text{Int}(\boldsymbol{A})$ and $\lim_{m\to\infty}
\Delta_{[m\mathbf{x}]}^{(m)}= 0$ if $\mathbf{x}\in
\boldsymbol{C}\setminus\boldsymbol{A}$, and the fact that
$\lim_{m\to\infty}\frac{1}{m} [m\mathbf{x}]= \mathbf{x}$.
\end{proof}

Let us fix a $\boldsymbol{\mu}\in\mathcal{P}^{(\infty )}$ and pick
$m$ sufficiently large so as ensure that
$\boldsymbol{\mu}\in\mathcal{P}^{(m)}$. We denote by
$\boldsymbol{\xi}^{(m)}_{\boldsymbol{\mu}}$ and
$\boldsymbol{\xi}^{(\infty)}_{\boldsymbol{\mu}}$ the associated
Bethe vectors detailed in Theorem \ref{bethe-sol:thm} and Theorem
\ref{yang-yang:thm}, respectively.
\begin{lemma}\label{speclim:lem}
For any $\boldsymbol{\mu}\in\mathcal{P}^{(\infty )}$, one has that
\begin{equation}
\lim_{m\to\infty}
m\boldsymbol{\xi}_{\boldsymbol{\mu}}^{(m)}=
\boldsymbol{\xi}_{\boldsymbol{\mu}}^{(\infty
)} .
\end{equation}
\end{lemma}
\begin{proof}
Let $m_j:=\langle
\boldsymbol{\mu},\boldsymbol{\alpha}_j\rangle+\cdots +\langle
\boldsymbol{\mu},\boldsymbol{\alpha}_{n-1}\rangle +n-j$, $j=1,\ldots
,n$. Then $\boldsymbol{\xi}_{\boldsymbol{\mu}}^{(m)}$ and
$\boldsymbol{\xi}_{\boldsymbol{\mu}}^{(\infty)}$ correspond to (the
projections onto the center-of-mass plane of) the (unique) global
minima of $V(\xi_1,\ldots ,\xi_n)$ \eqref{V} and
$V^{(\infty)}(\xi_1,\ldots ,\xi_n)$ \eqref{Vinf}, respectively. The
rescaled  Bethe vector $m\boldsymbol{\xi}_{\boldsymbol{\mu}}^{(m)}$
thus corresponds to the global minimum of the function
$V^{(m)}(\xi_1,\ldots ,\xi_n):=m V(\xi_1/m,\ldots ,\xi_n/m)$. The
lemma now follows from the observation that for $m\to\infty$ the
strictly convex function $V^{(m)}(\xi_1,\ldots ,\xi_n)$ tends to
$V^{(\infty)}(\xi_1,\ldots ,\xi_n)$ uniformly on compacts (which
implies in particular that the global minimum of the $V^{(m)}$
converges to the global minimum of $V^{(\infty)}$).
\end{proof}

The proof of the orthogonality in Theorem \ref{dorlas:thm} now
hinges on the following proposition.

\begin{proposition}\label{lim:prp}
For all $\boldsymbol{\mu}, \boldsymbol{\mu}^\prime\in
\mathcal{P}^{(\infty )}$, one has that
\begin{eqnarray}
\lefteqn{\lim_{m\to\infty} \int_{\boldsymbol{C}} \Psi^{(m)}
(\mathbf{x},\boldsymbol{\xi}^{(m)}_{\boldsymbol{\mu}}) \overline{
\Psi^{(m)}
(\mathbf{x},\boldsymbol{\xi}^{(m)}_{\boldsymbol{\mu}^\prime})}\text{d}\mathbf{x}}
&& \\
&& = \int_{\boldsymbol{A}} \Psi^{(\infty)}
(\mathbf{x},\boldsymbol{\xi}^{(\infty)}_{\boldsymbol{\mu}})
\overline{\Psi^{(\infty)}
(\mathbf{x},\boldsymbol{\xi}^{(\infty)}_{\boldsymbol{\mu}^\prime})}
\text{d}\mathbf{x}
.\nonumber
\end{eqnarray}
\end{proposition}
\begin{proof}
It is clear from (the proof of) Lemma \ref{wflim:lem} and from Lemma
\ref{speclim:lem} that the integrand and support of the integral on
the l.h.s. converges pointwise to the integrand and support of the
integral on the r.h.s. To see that the integrals themselves converge
accordingly we write
\begin{eqnarray*}
\lefteqn{ \int_{\boldsymbol{C}} \Psi^{(m)}
(\mathbf{x},\boldsymbol{\xi}) \overline{ \Psi^{(m)}
(\mathbf{x},\boldsymbol{\xi}^\prime})\text{d}\mathbf{x}}
&& \\
&& = \sum_{\sigma,\sigma^\prime\in\mathcal{S}_n}
\hat{\mathcal{C}}(\boldsymbol{\xi}_\sigma)
\hat{\mathcal{C}}(-\boldsymbol{\xi}^\prime_{\sigma^\prime})
\int_{(1+\frac{n}{m})\boldsymbol{A}} e^{i\langle [m\mathbf{x}]
,\boldsymbol{\xi}_\sigma
-\boldsymbol{\xi}^\prime_{\sigma^\prime}\rangle}
\Delta^{(m)}_{[m\mathbf{x}]} \text{d}\mathbf{x} ,
\end{eqnarray*}
where
$\hat{\mathcal{C}}(\boldsymbol{\xi})=\prod_{\boldsymbol{\alpha}\in
\mathbf{R}^+} \frac{1-e^{-g/m}\, e^{-i\langle
\boldsymbol{\alpha},\boldsymbol{\xi}\rangle} }{1-e^{-i\langle
\boldsymbol{\alpha},\boldsymbol{\xi}\rangle} }$. After substituting
$\boldsymbol{\xi}:=\boldsymbol{\xi}^{(m)}_{\boldsymbol{\mu}}$ and
$\boldsymbol{\xi}^\prime:=\boldsymbol{\xi}^{(m)}_{\boldsymbol{\mu}^\prime}$
the proposition follows for $m\to \infty$ upon invoking Lemma
\ref{speclim:lem} and the dominated convergence theorem of Lebesgue.
Indeed, one has that
\begin{equation*}
e^{i\frac{1}{m}\langle [m\mathbf{x}]
,m\sigma(\boldsymbol{\xi}^{(m)}_{\boldsymbol{\mu}}) -m\sigma^\prime
(\boldsymbol{\xi}^{(m)}_{\boldsymbol{\mu}^\prime})\rangle}\longrightarrow
e^{i\langle \mathbf{x} ,\sigma(\boldsymbol{\xi}^{(\infty
)}_{\boldsymbol{\mu}}) -\sigma^\prime(\boldsymbol{\xi}^{(\infty
)}_{\boldsymbol{\mu}^\prime})\rangle}
\end{equation*}
and
\begin{equation*}
\Delta_{[m\mathbf{x}]}^{(m)}\longrightarrow
\begin{cases}
1 & \text{if}\  \mathbf{x}\in\text{Int}(\boldsymbol{A}) \\
0 &\text{if}\ \mathbf{x}\in \boldsymbol{C}\setminus\boldsymbol{A}
\end{cases}
\end{equation*}
pointwise for $m\to \infty$, and that $| e^{i\frac{1}{m}\langle
[m\mathbf{x}] ,m\sigma(\boldsymbol{\xi}^{(m)}_{\boldsymbol{\mu}})
-m\sigma^\prime
(\boldsymbol{\xi}^{(m)}_{\boldsymbol{\mu}^\prime})\rangle} |=1$,
$|\Delta_{[m\mathbf{x}]}^{(m)}|\leq 1$.
\end{proof}

Proposition \ref{lim:prp} can be rephrased as
\begin{eqnarray*}
\lefteqn{\langle \Psi^{(\infty)}
(\boldsymbol{\xi}^{(\infty)}_{\boldsymbol{\mu}}), \Psi^{(\infty)}
(\boldsymbol{\xi}^{(\infty)}_{\boldsymbol{\mu}^\prime})\rangle^{(\infty
)}=} && \\
&& \lim_{m\to\infty} \int_{\boldsymbol{C}} (J^{(m)}\Psi
(\boldsymbol{\xi}^{(m)}_{\boldsymbol{\mu}}))(\mathbf{x})\overline{
(J^{(m)}\Psi (\boldsymbol{\xi}^{(m)}_{\boldsymbol{\mu}^\prime}))
(\mathbf{x})}\text{d}\mathbf{x} .
\end{eqnarray*}
The r.h.s. of this limiting relation vanishes when
$\boldsymbol{\mu}\neq \boldsymbol{\mu}^\prime $ in view of Eq.
\eqref{ipr} and Theorem \ref{oc:thm}, whence the orthogonality in
Theorem \ref{dorlas:thm} follows.

\subsection{Hamiltonian}
We will now wrap up by verifying briefly that formally the
Hamiltonian $H^{(m)}$ \eqref{ham1}, \eqref{ham2} converges in the
continuum limit to the Hamiltonian of the repulsive delta Bose gas
on the circle. It is quite plausible that with a somewhat more
in-depth analysis in the spirit of Ref. \cite{rui:continuum} one
would be able to show that this convergence of the Hamiltonian is in
fact in the strong resolvent sense, but we will not attempt to do so
here.

Let $\textsc{H}^{(\infty )}$ be the self-adjoint extension in
$\mathcal{H}^{(\infty)}=L^2(\boldsymbol{A},\text{d}\mathbf{x})$ of
the Laplace operator $-\Delta $ with boundary conditions of the from
in Eqs. \eqref{ba}, \eqref{bb}, and let $\textsc{H}^{(m)}$ be the
following rescaled staircase embedding of the operator $H^{(m)}$
\eqref{ham1}--\eqref{ham2} in
$L^2(\boldsymbol{C},\text{d}\mathbf{x})$:
\begin{equation}
\textsc{H}^{(m)}=2 m^2 J^{(m)} H^{(m)} (J^{(m)})^{-1} \Pi^{(m)},
\end{equation}
where $\Pi^{(m)}:L^2(\boldsymbol{C},\text{d}\mathbf{x})\to
L^2(\boldsymbol{C},\text{d}\mathbf{x})$ denotes the orthogonal
projection onto the finite-dimensional subspace of staircase
functions $J^{(m)}(\mathcal{H}^{(m)})\subset
L^2(\boldsymbol{C},\text{d}\mathbf{x})$. It is clear that
\begin{subequations}
\begin{equation}\label{eveqinf}
\textsc{H}^{(\infty )} \Psi^{(\infty
)}(\boldsymbol{\xi}_{\boldsymbol{\mu}}^{(\infty)})=
E^{(\infty)}(\boldsymbol{\xi}_{\boldsymbol{\mu}}^{(\infty)})
\Psi^{(\infty )}(\boldsymbol{\xi}_{\boldsymbol{\mu}}^{(\infty)}),
\end{equation}
with $E^{(\infty)}(\boldsymbol{\xi})=\langle
\boldsymbol{\xi},\boldsymbol{\xi}\rangle$, and that
\begin{equation}\label{eveqm}
\textsc{H}^{(m )} \Psi^{(m
)}(\boldsymbol{\xi}_{\boldsymbol{\mu}}^{(m)})=
E^{(m)}(\boldsymbol{\xi}_{\boldsymbol{\mu}}^{(m)}) \Psi^{(m
)}(\boldsymbol{\xi}_{\boldsymbol{\mu}}^{(m)}),
\end{equation}
\end{subequations}
where $E^{(m)}(\boldsymbol{\xi}):=2m^2 E(\boldsymbol{\xi})$ with
$E(\boldsymbol{\xi} )$ given by Eq. \eqref{evham}. From Lemmas
\ref{wflim:lem} and \ref{speclim:lem} it follows that
$\lim_{m\to\infty}\Psi^{(m
)}(\mathbf{x},\boldsymbol{\xi}_{\boldsymbol{\mu}}^{(m)})=
\Psi^{(\infty
)}(\mathbf{x},\boldsymbol{\xi}_{\boldsymbol{\mu}}^{(\infty)})$
pointwise for $\mathbf{x}\in\text{Int}(\boldsymbol{A})$ and that
$\lim_{m\to\infty}
E^{(m)}(\boldsymbol{\xi}_{\boldsymbol{\mu}}^{(m)})=
E^{(\infty)}(\boldsymbol{\xi}_{\boldsymbol{\mu}}^{(\infty)})$.
In other words, for $m\to\infty$  the eigenfunctions, the
eigenvalues, and the eigenvalue equation for $\textsc{H}^{(m )}$ in
Eq. \eqref{eveqm} converge pointwise to the eigenfunctions, the
eigenvalues, and the eigenvalue equation for $\textsc{H}^{(\infty
)}$ in Eq. \eqref{eveqinf}, respectively.

\vspace{3ex}

{\bf Acknowledgments.} Thanks are due to M. Bustamante and to S.N.M.
Ruijsenaars for several helpful discussions.

\bibliographystyle{amsplain}

\end{document}